\documentclass[12pt]{article}
\usepackage{amsmath,amssymb,amsfonts,latexsym}
\usepackage{color}
\usepackage{graphicx}

\catcode `\@=11 \@addtoreset{equation}{section}
\def\theequation{\arabic{section}.\arabic{equation}}
\catcode `\@=12

\newcommand{\be}{\begin{equation}}
\newcommand{\en}{\end{equation}}
\newcommand{\bea}{\begin{eqnarray}}
\newcommand{\ena}{\end{eqnarray}}
\newcommand{\beano}{\begin{eqnarray*}}
\newcommand{\enano}{\end{eqnarray*}}
\newcommand{\bee}{\begin{enumerate}}
\newcommand{\ene}{\end{enumerate}}
\newcommand{\R}{\mathcal{R}}

\newcommand{\ST}{\mathcal{S}}

\newcommand{\Hil}{\mathcal{H}}

\newcommand{\Id}{1\!\!1}

\newcommand{\K}{\mathcal{K}}

\newcommand{\Sc}{\mathcal{S}}

\newcommand{\M}{\mathcal{M}}
\newcommand{\A}{\mathcal{A}}

\newcommand{\1}{1 \!\!\! 1}

\newtheorem{rem}{Remark}

\begin{document}
\title{A phenomenological operator description of interactions between populations with applications to migration}

\author{{\Large Fabio Bagarello}\\
{\small Dieetcam}\\
{\small Facolt\`a di Ingegneria, Universit\`a di Palermo}\\
{\small Viale delle Scienze, I--90128  Palermo, Italy}\\
{\small e--mail: bagarell@unipa.it}\\
\vspace{2mm}\\
{\Large Francesco Oliveri}\\
{\small Dipartimento di Matematica, Universit\`a di Messina}\\
{\small Viale F. Stagno D'Alcontres 31, I--98166 Messina, Italy}\\
{\small e--mail: foliveri@unime.it}}

\date{}
\maketitle
\begin{abstract}
\noindent We adopt an operatorial method based on the so--called creation, annihilation and number operators in the description of different
systems in which two populations interact and move in a two--dimensional region. In particular, we discuss diffusion processes modeled by a
quadratic hamiltonian. This general procedure will be adopted, in particular, in the description of migration phenomena. With respect to our
previous analogous results,  we use here fermionic operators since they automatically implement an upper bound for the population densities.
\end{abstract}

\vspace{1cm}



\newpage

\section{Introduction and preliminaries}
\label{sec:introduction}

A large body of theoretical and experimental evidence that spatial patchy environments influence the dynamics of species interactions is
available in the literature \cite{hanski1981,hanski1983,ives-may,comins-hassel,comins-hassel-may,hanski2008}. Hence, a broad variety of
spatially extended models has been developed in theoretical biology. The principle of competitive exclusion (Gause's Law), stating that two
like (identical) species cannot coexist in the same habitat, is violated in patchy environments where two like species may coexist due to
migration \cite{Slatkin 1974}. A lot of evidence exists about the coexistence, as metapopulations \cite{Hanski1991} in a fragmented
environment, of two competing species (or populations)  even if one is competitively superior to the other; in fact, the local extinction in a
patch of the inferior competitor has no global effect if it is able to disperse more effectively into unoccupied patches (see
\cite{Nee-May1992} and the references therein quoted).

Besides the usual models based on continuous reaction--diffusion equations \cite{murray} and cellular automata, the coupled map lattice (CML)
formalism has been widely used in the simulation of biological spatial interactions. In the usual CML approach, local prey--predator (or
host--parassitoid) dynamics are coupled with their $n$--nearest neighbors through some appropriate exchange rule. Populations interact and
disperse over the points of a lattice (used to simulate the patchy environment). In such a context various aspects can be observed, such as the
emergence of some persistent spatial patterns in the distributions of the competing species (\emph{e.g.}, phytoplankton distribution in the
oceans), or the phenomena of synchronization between the phases of nearby regions \cite{PaparellaOliveri2008}.

The mathematical oriented literature on migration is very poor. But for those cited above (more concerned with the coexistence of biological
species), only few papers deal with this problem. For instance, in \cite{quint}, the authors propose a game-theoretic model of migration on
animals, while in \cite{Bij}, the author concentrates his attention to migration in The Netherlands. Moreover, it is worth of note the paper \cite{zhang} where an oscillating behavior has been described with Sheppard's generalized dynamic migration model.

Other models of competing populations including also spatial interactions can be constructed using completely different tools, say, operator
methods of quantum mechanics. In a series of recent papers one of us (F.B.) used some tools from quantum mechanics, like operator algebras and,
in particular, the so--called \emph{number representation}, to discuss some toy models of stock markets \cite{bag1,bag2,bag3,bag4}. More
recently, we have adopted the same framework in the analysis of {\em love affairs}, \cite{ff1,ff2,bag5}. In these rather different areas the
use of the number representation proved to be quite useful to set up a natural description of the system under consideration. In fact, several
quantities which change discontinuously are very well described in terms of the integer eigenvalues of certain relevant self--adjoint
operators, the {\em observables} of the system, and the dynamics is described by an energy--like operator, the {\em hamiltonian}. This approach
has produced several interesting results like, just to cite a few, the possibility of describing the time evolution of the portfolio of the
traders of a simplified stock markets, or of computing the transition probability of the market itself from a given initial state ({\em i.e.},
from a given distribution of the cash and the shares for the different traders) to a final state. In the context of love affairs, we have found
a non--trivial dynamical behavior of the actors of the affairs (Alice and Bob), even in the presence of a third actress, Carla, and we have
also considered the effect of a {\em reservoir} interacting with them, mimicking the real world.

In this paper, we use an analogous strategy, based on fermionic  rather than bosonic operators, to describe some models arising in different
contexts. In particular, we will describe a (strictly local) situation in which two populations live together and are forced to interact, and a
second situation in which the two species occupy (in general) different cells of a two--dimensional lattice, interact and move along the cells.
These models can be useful to model many different systems. In particular, we will restrict ourselves to consider the following two: a
migration process in which a population moves from a given (poor) place to a richer region of the lattice, which is already occupied by a
second group of people, and a  system in which the two populations compete.

The main motivations which suggest the use of the tools originally developed in a quantum context to describe classical situations have been
widely discussed along the years, \cite{bag1,bag2,bag3,bag4,ff1,ff2,bag5}, and will not be repeated here. We just want to mention that, in
recent years, the intersection between {\em quantum} and {\em classical} worlds has became more and more frequent. We refer to
\cite{qal,khre,baa}, and to the references therein, for many other applications and for some general comments.

The choice of using  fermionic operators rather than bosonic ones is mainly based on two reasons. The first one is of technical nature: the
Hilbert space of our models is automatically finite dimensional, so that all the observables are bounded operators.  The second reason is
related to the biological and/or sociological interpretation of our model: for each population which will be considered, we will have two only
possible non--trivial situations. In the first one (the {\em ground state}) there is a very low density, while in the second one (the {\em
excited state}), the density is very high. Hence, if we try to increase the density of the excited state, or if we try to decrease the density
of the ground state, we simply annihilate that population! We can interpret this fact just saying that there exists upper and lower bounds to
the densities of the populations which can not be overcame for obvious reasons: for instance, because the environment can not give enough food
to the populations. Of course, this rather sharp division in just two levels may appear unsatisfactory. However, it is not hard to extend our
procedure to an arbitrary number of levels, paying the price of some technical difficulties. We will not consider this extension here since,
already in our hypotheses, an absolutely non--trivial and realistic dynamics can be still deduced.


The paper is organized as follows. In Section~\ref{sec:firstmodel}, we consider a first simple model involving two populations, and we analyze
the dynamics of their relationship starting from  very natural assumptions. The considered model is linear, strictly local, and the equations
of motion are solved analytically.

In Section~\ref{sec:spreading}, we extend this  model by allowing a spatial distribution. The interaction is quadratic, so that the solution
can be again deduced analytically.  The model will be discussed in terms of migrant and resident populations.

In Section~\ref{sec:spreading2}, we show how the same model introduced in Section~\ref{sec:spreading}, with a different choice of the
parameters and of the initial conditions, can be used in the description of two competing populations.

Finally, Section~\ref{sec:conclusions} contains our conclusions, while, to keep the paper self--contained, we list in the Appendix few useful
facts about quantum mechanics and the number representation for fermions. In all the considered cases a quasi--periodic dynamics is obtained
and the two populations coexist in the patchy environment.

\section{A first model}
\label{sec:firstmodel} In this section, we introduce a first simple model which is useful to fix the main ideas and the notation. This model
will also be used in the next section as a building block of a more sophisticated model. In particular,
no spatial distribution will be
considered. Following \cite{bag1,bag2,bag3,bag4,ff1,ff2,bag5}, we associate to each population $\Sc_j$ of our model an annihilation and a
creation operator $a_j$ and $a_j^\dagger$, and a related number operator $\hat n_j:=a_j^\dagger a_j$. Here, we just consider two populations,
$\Sc_1$ and $\Sc_2$. We assume the following anticommutation rules: \be \{a_i,a_j^\dagger\}=\delta_{i,j},\qquad
\{a_i,a_j\}=\{a_i^\dagger,a_j^\dagger\}=0, \label{21}\en $i,j=1,2$. Recall that $\{x,y\}=xy+yx$. These rules imply in particular that
$a_j^2=\left(a_j^\dagger\right)^2=0$ (see the Appendix). Hence, if $\varphi_{0,0}$ is the ground state, $a_1\varphi_{0,0}=a_2\varphi_{0,0}=0$,
the only non--trivial vectors of our Hilbert space $\Hil$ are
\[
\varphi_{0,0},\qquad \varphi_{1,0}:=a_1^\dagger\varphi_{0,0}, \qquad \varphi_{0,1}:=a_2^\dagger\varphi_{0,0}, \qquad \varphi_{1,1}:=a_1^\dagger a_2^\dagger\varphi_{0,0}.
\]
This means that $dim(\Hil)=4$. The biological interpretation of these vectors follows from the following eigenvalue equations: \be \hat
n_1\varphi_{n_1,n_2}=n_1\varphi_{n_1,n_2},\qquad \hat n_2\varphi_{n_1,n_2}=n_2\varphi_{n_1,n_2}, \label{22} \en $j=1,2$, where $\hat
n_j=a_j^\dagger a_j$ is the number operator of $\Sc_j$ (see the Appendix). Saying that $\varphi_{0,0}$ is {\em the state of the system} means
that there are very few subjects of the two populations in our region. If the state is $\varphi_{1,0}$, then there are very few elements of
$\Sc_2$ but very many elements of $\Sc_1$. The opposite situation is described by $\varphi_{0,1}$, while $\varphi_{1,1}$ describes the case in
which both populations are abundant. As already stated, it is not possible to have, {\em e.g.}, more elements of $\Sc_1$ than those described
by $\varphi_{1,0}$ or $\varphi_{1,1}$: trying to further increase the density of $\Sc_1$ simply destroys this population! This is a simple
consequence of $(a_1^\dagger)^2=0$. As in our previous applications, we use a self--adjoint operator, which we call the {\em hamiltonian} of
the system, to derive the dynamics of the system, and which describes the interaction between the populations. The self--adjoint hamiltonian
which we assume here is the following one: \be H=H_0+\lambda H_I,\qquad H_0=\omega_1a_1^\dagger a_1+\omega_2a_2^\dagger a_2, \quad
H_I=a_1^\dagger a_2+a_2^\dagger a_1, \label{23}\en in which $\omega_j$ and $\lambda$ are real positive quantities. In particular, it is
$\lambda=0$ when the two populations do not interact. In this case, $H$ describes a static situation, in which the densities of the two
populations, described by the number operators $\hat n_j$, do not change with $t$. This is a consequence of the fact that $[H_0,\hat n_j]=0$,
$j=1,2$. On the other hand, if $\lambda\neq0$, then $H_I$ describes a situation in which the density of $\Sc_1$ increases (because of
$a_1^\dagger$) while that of $\Sc_2$ (decreases because of $a_2$), or viceversa, \cite{bagijtp}. The equations of motion that are obtained are
(see (\ref{a2}) in the Appendix): \be
\begin{aligned}
&\dot a_1(t)=-i\omega_1 a_1(t)-i\lambda a_2(t),\\
&\dot a_2(t)=-i\omega_2 a_2(t)-i\lambda a_1(t),
\end{aligned}
\label{24} \en which can be solved with the initial conditions $a_1(0)=a_1$ and $a_2(0)=a_2$. The solution looks like \be
\begin{aligned}
&a_1(t)=\frac{1}{2\delta}\left(a_1\left((\omega_1-\omega_2)\Phi_-(t)+\delta\Phi_+(t)\right)+2\lambda a_2\Phi_-(t)\right),\\
&a_2(t)=\frac{1}{2\delta}\left(a_2\left(-(\omega_1-\omega_2)\Phi_-(t)+\delta\Phi_+(t)\right)+2\lambda a_1\Phi_-(t)\right),
\end{aligned}
\label{25} \en where
\[
\begin{aligned}
&\delta=\sqrt{(\omega_1-\omega_2)^2+4\lambda^2},\\
&\Phi_+(t)=2\exp\left(-\frac{it(\omega_1+\omega_2)}{2}\right)\cos\left(\frac{\delta t}{2}\right),\\
&\Phi_-(t)=-2i\exp\left(-\frac{it(\omega_1+\omega_2)}{2}\right)\sin\left(\frac{\delta t}{2}\right).
\end{aligned}
\]
It is now easy to deduce the mean value of the time evolution of the number operator $\hat n_j(t)$, which, as discussed before, we interpret
here as  {\em the density of $\Sc_j$}: $n_j(t):=\left<\varphi_{n_1,n_2}, \hat n_j(t)\varphi_{n_1,n_2}\right>$. More explicitly, $n_j(t)$ is the
time evolution of the density of $\Sc_j$ assuming that, at $t=0$, the density of $\Sc_1$ was $n_1$ and that of $\Sc_2$ was $n_2$, the quantum
numbers labeling the state $\varphi_{n_1,n_2}$.  Using (\ref{22}) and the orthonormality of the different $\varphi_{n1,n2}$'s, we obtain \be
n_1(t)=n_1\frac{(\omega_1-\omega_2)^2}{(\omega_1-\omega_2)^2+4\lambda^2}+ \frac{4\lambda^2}{(\omega_1-\omega_2)^2+4\lambda^2}
\left\{n_1\cos^2\left(\frac{\delta t}{2}\right)+n_2\sin^2\left(\frac{\delta t}{2}\right)\right\}, \label{26}\en and \be
n_2(t)=n_2\frac{(\omega_1-\omega_2)^2}{(\omega_1-\omega_2)^2+4\lambda^2}+ \frac{4\lambda^2}{(\omega_1-\omega_2)^2+4\lambda^2}
\left\{n_2\cos^2\left(\frac{\delta t}{2}\right)+n_1\sin^2\left(\frac{\delta t}{2}\right)\right\}. \label{27}\en Notice that these formulas
imply that $n_1(t)+n_2(t)=n_1+n_2$, independently of $t$ and $\lambda$. This is expected, since it is easy to check that $[H,\hat n_1+\hat
n_2]=0$. Secondly, since $n_1$ and $n_2$ can only be 0 or 1, we can also check that, if $n_1=n_2=n$, then $n_1(t)=n_2(t)=n$ for all $t$: if the
two populations are equally distributed at $t=0$, then they do not change their distributions. If $n_1=1$ and $n_2=0$, then
\[
n_1(t)=1-\frac{4\lambda^2}{\delta^2}\sin^2\left(\frac{t\delta}{2}\right), \qquad
n_2(t)=\frac{4\lambda^2}{\delta^2}\sin^2\left(\frac{t\delta}{2}\right),
\]
while if $n_1=0$ and $n_2=1$, then
\[
n_2(t)=1-\frac{4\lambda^2}{\delta^2}\sin^2\left(\frac{t\delta}{2}\right), \qquad
n_1(t)=\frac{4\lambda^2}{\delta^2}\sin^2\left(\frac{t\delta}{2}\right).
\]
In all these cases we have $0\leq n_j(t)\leq 1$ for all $t$, as it should be.

Since $n_1(t)+n_2(t)=n_1+n_2$, we find that $\delta n(t):=|n_1(t)-n_1|=|n_2(t)-n_2|$ which, in the two cases above, gives
\[
\delta n(t)=\frac{4\lambda^2}{\delta^2}\,\sin^2\left(\frac{t\delta}{2}\right),
\]
so that the variations of the two populations coincide. In general, equation
(\ref{26}) gives
\[
n_1(t)-n_1=\frac{4\lambda^2}{\delta^2}\,(n_2-n_1)\,\sin^2\left(\frac{t\delta}{2}\right),
\]
which is in agreement with the
previous result since $n_1$ and $n_2$ can only be 0 or 1. Notice also that, if $n_1=n_2$, then $n_1(t)=n_1$ for all $t$, as already remarked.

Now let us restrict ourselves, for concreteness, to the case $n_1=1$ and $n_2=0$. Hence
\[
\Delta_n:=\max \{\delta
n(t)\}=\frac{1}{1+\left(\frac{\omega_1-\omega_2}{2\lambda}\right)^2}, \qquad
\hbox{for} \; \frac{t\delta}{2}=\frac{\pi}{2},\frac{3\pi}{2},\ldots,
\]
while
\[
\min
\{\delta n(t)\}=0\qquad
\hbox{for} \; \frac{t\delta}{2}=0,\pi,2\pi,\ldots.
\]

In particular, $\Delta_n$  is almost equal to 1  if $\omega_1\simeq\omega_2$,
independently of $\lambda\neq0$, while is almost 0, as $\lambda$ is kept fixed, when $|\omega_1-\omega_2|$ is very large. Incidentally, if
$\lambda=0$ the two species do not interact and, in fact, $\Delta_n=0$: the model has essentially no dynamics. For obvious reasons we are only
interested to the case $\lambda>0$. As already mentioned before, these results show that the free hamiltonian, which does not affect the density of the
populations if $\lambda=0$, produces a non--trivial effect if $\lambda\neq0$. A similar conclusion will be deduced in Sections
\ref{sec:spreading} and \ref{sec:spreading2}, where we consider the spatial version of this model. More in details, $|\omega_1-\omega_2|$ can be
considered as a sort of {\em inertia} of the system: the larger its value is, the smaller the variations of $n_j(t)-n_j$ are. On the other hand,
if $|\omega_1-\omega_2|\simeq 0$, then the system has a very low inertia and, in fact, very large changes in the densities of both
populations are allowed. An interesting fact is that  only the difference between the two frequencies $\omega_1$ and $\omega_2$ play a role in
the dynamics of both $\Sc_1$ and $\Sc_2$.

Concerning a relation between the $\omega$'s and $\lambda$, from (\ref{26}) and (\ref{27}) we  also see that if $|\omega_1-\omega_2|\gg
2\lambda$ there is essentially no dynamics: $n_j(t)\simeq n_j$, $j=1,2$. On the contrary, if $|\omega_1-\omega_2|\ll 2\lambda$, the constant
contributions in  (\ref{26}) and (\ref{27}) are very small compared with the oscillating contributions. These results agree with our previous
conclusions.

\section{A spatial model}
\label{sec:spreading}

In this section, we extend the model  introduced above with the aim of including spatial effects: we consider a 2D--region $\R$ in which, in
principle, the two populations are distributed. Under reasonable assumptions, a simple model for $\Sc_1$ and $\Sc_2$ can be deduced, and its
dynamics investigated.

The starting point is the ({\em e.g.}, rectangular or square) region $\R$, which we divide in $N$ cells, labeled by $\alpha=1,2,\ldots,N=L\cdot
L^\prime$. With $\alpha=1$ we label the first cell down to the left, while $N$ is the last cell, up to the right (see Figure~\ref{fig1}).

\begin{figure}
\begin{center}
\begin{picture}(400,180)
\put(37,180){\thicklines\line(1,0){330}}
\put(37,-30){\thicklines\line(0,1){210}}
\put(37,-30){\thicklines\line(1,0){330}}
\put(367,-30){\thicklines\line(0,1){210}}
\put(37,0){\line(1,0){330}}
\put(37,30){\line(1,0){330}}
\put(37,60){\line(1,0){330}}
\put(37,90){\line(1,0){330}}
\put(37,120){\line(1,0){330}}
\put(37,150){\line(1,0){330}}
\put(67,-30){\line(0,1){210}}
\put(97,-30){\line(0,1){210}}
\put(127,-30){\line(0,1){210}}
\put(157,-30){\line(0,1){210}}
\put(187,-30){\line(0,1){210}}
\put(217,-30){\line(0,1){210}}
\put(247,-30){\line(0,1){210}}
\put(277,-30){\line(0,1){210}}
\put(307,-30){\line(0,1){210}}
\put(337,-30){\line(0,1){210}}
\put(53,-15){\makebox(0,0){\footnotesize$1$}}
\put(80,-15){\makebox(0,0){\footnotesize$2$}}
\put(110,-15){\makebox(0,0){\footnotesize$\dots$}}
\put(140,-15){\makebox(0,0){\footnotesize$\dots$}}
\put(290,-15){\makebox(0,0){\footnotesize$\dots$}}
\put(322,-15){\makebox(0,0){\footnotesize$L-1$}}
\put(353,-15){\makebox(0,0){\footnotesize$L$}}
\put(52,15){\makebox(0,0){\footnotesize$L+1$}}
\put(82,15){\makebox(0,0){\footnotesize$L+2$}}
\put(110,15){\makebox(0,0){\footnotesize$\cdots$}}
\put(140,15){\makebox(0,0){\footnotesize$\cdots$}}
\put(322,165){\makebox(0,0){\footnotesize$\cdots$}}
\put(352,165){\makebox(0,0){\footnotesize$L\cdot L^\prime$}}
\put(53,165){\makebox(0,0){\footnotesize$L'$}}
\end{picture}
\end{center}
\vspace*{5mm}
\caption{\label{fig1} The two--dimensional lattice for the spatial model.}
\end{figure}
\vspace*{.1cm}

The main idea of our model here is that in each cell $\alpha$ the two populations, whose related operators are $a_\alpha$, $a_\alpha^\dagger$
and $\hat n^{(a)}_\alpha=a_\alpha^\dagger a_\alpha$ for what concerns $\Sc_1$, and $b_\alpha$, $b_\alpha^\dagger$ and $\hat
n^{(b)}_\alpha=b_\alpha^\dagger b_\alpha$ for $\Sc_2$, behave as in the previous section. This means that the same hamiltonian as in (\ref{23})
will be assumed here in $\alpha$. Using our new notation, we define \be H_\alpha=H_\alpha^0+\lambda_\alpha H_\alpha^I,\qquad
H_\alpha^0=\omega_\alpha^a a_\alpha^\dagger a_\alpha+\omega_\alpha^b b_\alpha^\dagger b_\alpha, \quad H_\alpha^I=a_\alpha^\dagger
b_\alpha+b_\alpha^\dagger a_\alpha. \label{31} \en Extending what we have discussed in the previous section, it is natural to interpret the
mean values of the operators $\hat n^{(a)}_\alpha$ and $\hat n^{(b)}_\alpha$ as {\em local density operators} (the local densities are in the
sense of mixtures; hence, we may sum up local densities relative to different cells) of the two populations in the cell $\alpha$: if the mean
value of, say, $\hat n^{(a)}_\alpha$, in the state of the system is equal to one, this means that the density of $\Sc_1$ in the cell $\alpha$
is very high. Notice that $H_\alpha=H_\alpha^\dagger$, since all the parameters, which in general are assumed to be cell--depending (to allow
for the description of an anisotropic situation), are real and positive numbers. The anticommutation rules are those in (\ref{21}), which we
rewrite as \be \{a_\alpha,a_\beta^\dagger\}=\{b_\alpha,b_\beta^\dagger\}=\delta_{\alpha,\beta}, \qquad \{a_\alpha^\sharp,b_\beta^\sharp\}=0.
\label{32} \en Of course, the hamiltonian $H$ must consist of a sum of all the different $H_\alpha$ plus another contribution, $h$, responsible
for the diffusion of the populations all around the lattice. A natural choice for $h$ is the following one: \be
h=\sum_{\alpha,\beta}p_{\alpha,\beta}\left\{\gamma_a\left(a_\alpha a_\beta^\dagger+a_\beta a_\alpha^\dagger\right)+\gamma_b\left(b_\alpha
b_\beta^\dagger+b_\beta b_\alpha^\dagger\right)\right\}, \label{33} \en where also $\gamma_a$, $\gamma_b$ and the $p_{\alpha,\beta}$ are real
quantities. In particular, $p_{\alpha,\beta}$ can only be 0 or 1 depending on the possibility of the populations to move from cell $\alpha$ to
cell $\beta$ or vice-versa. In fact, this is the meaning of, e.g., the contribution $a_\alpha a_\beta^\dagger$ in (\ref{33}), \cite{bagijtp}:
because of $a_\alpha$ the mean value of $\hat n^{(a)}_\alpha$ decreases, while the mean value of $\hat n^{(a)}_\beta$ increases because of
$a_\beta^\dagger$. We interpret this feature saying that some individuals of $\Sc_1$ are moving from the cell $\alpha$ to the cell $\beta$, if
$p_{\alpha,\beta}=1$. For this reason the $p_{\alpha,\beta}$'s can be considered as {\em diffusion coefficients}. Notice that a similar role is
also played by $\gamma_a$ and $\gamma_b$, which, however, could assume different values. This is important since, in this way, a different
mobility for the two species can be introduced. In the rest of the paper we will assume that diffusion may take place only between nearest
neighboring cells. Of course, we should clarify what we mean by {\em neighboring}: in this paper, we will consider a simple planar topology, in
which the neighboring cells of the cell labeled $\alpha$ are the cells  $\alpha-1$, $\alpha+1$, $\alpha+L$ and $\alpha-L$, provided that they
exist (this is verified for the internal cells of the lattice); for the cells located along the boundaries of the lattice we have only three
neighbors (two neighbors for the four cells located at the corners of the lattice). This is a natural choice for the physical system we have in
mind here. However, different choices could also be considered. For instance, we could also use a torus topology in which all the cells have
four neighboring cells. To deal with  these different topologies, it is enough to modify the diffusion coefficients. We will assume that
$p_{\alpha,\alpha}=0$ and that, for symmetry reasons, $p_{\alpha,\beta}=p_{\beta,\alpha}$. Now we define $H=\sum_\alpha H_\alpha+h$, which is
self-adjoint. The differential equations for the annihilation operators (see (\ref{a2}) in the Appendix) read \be
\begin{aligned}
&\dot a_\alpha=-i\omega_\alpha^a a_\alpha -i\lambda_\alpha b_\alpha +2i\gamma_a\sum_\beta p_{\alpha,\beta}a_\beta,\\
&\dot b_\alpha=-i\omega_\alpha^b b_\alpha -i\lambda_\alpha a_\alpha +2i\gamma_b\sum_\beta p_{\alpha,\beta}b_\beta.
\end{aligned}
\label{34} \en

\begin{rem}
Readers with a background in quantum many-body could interpret the operator $H$ as the hamiltonian of two kinds of fermions mutually
interacting. From this point of view, it could be interesting consider other aspects of the system, and not only its dynamics. Phase
transitions, Green's functions, non-zero temperature states are all typical problems which could be of a certain interest starting from a
similar hamiltonian, \cite{kit}.
\end{rem}

\begin{rem}
It might be worth noticing that the fermionic operators could be replaced by Pauli matrices. Using these operators (which again can be
considered as raising, lowering and number operators) we would obtain a formally different hamiltonian describing the same physics.
\end{rem}

\subsection{A simple case: equal coefficients}

As a first step, we suppose here that $\omega_\alpha^b=\omega_\alpha^a=\omega$, $\lambda_\alpha=\lambda$ and $\gamma_a=\gamma_b=\tilde\gamma$,
for all $\alpha\in\R$. Hence, by introducing $a_\alpha(t)=A_\alpha(t)e^{-i\omega t}$ and $b_\alpha(t)=B_\alpha(t)e^{-i\omega t}$, the above
equations can be easily rewritten as \be
\begin{aligned}
\dot A_\alpha= -i\lambda B_\alpha +2i\tilde\gamma\sum_\beta p_{\alpha,\beta}A_\beta,\\
\dot B_\alpha= -i\lambda B_\alpha +2i\tilde\gamma\sum_\beta p_{\alpha,\beta}B_\beta,
\end{aligned}
\label{35} \en which are true independently of the size of the region $\R$.

The diffusion coefficients $p_{\alpha,\beta}$  will all be zero but when $\alpha$ and $\beta$ refer to nearest neighbors in the planar
topology. In this case $p_{\alpha,\beta}=1$. We will consider now the situation of a square region $\R$ with $N=L^2$ cells, starting with the
simplest non--trivial situation, $L=2$. In this case, the only non zero diffusion coefficients are $p_{1,2}$, $p_{1,3},$ $p_{2,1},$ $p_{2,4},$
$p_{3,1}$, $p_{3,4}$, $p_{4,2}$ and $p_{4,3}$, all equal to one,  while the remaining ones are zero. Fixing natural initial conditions,
$A_\alpha(0)=a_\alpha$ and $B_\alpha(0)=b_\alpha$, we get, for instance,
\[
\begin{aligned}
A_1(t)&=\frac{1}{4}[2(a_1-a_4)\cos(t)+(a_1+a_4+b_2+b_3)\cos(7t)+(a_1+a_4-b_2-b_3)\cos(9t)\\
&-i(2(b_1-b_4+(b_1+b_4)\cos(8t))\sin(t)-(a_2+a_3)(\sin(7t)+\sin(9t)))],
\end{aligned}
\]
and so on. The number operators ({\em i.e.}, the local densities of $\Sc_1$ and $\Sc_2$) are deduced directly from the capital operators
$A_\gamma(t)$ and $B_\gamma(t)$, since $\hat n_\gamma^{(a)}(t)=a_\gamma^\dagger(t)a_\gamma(t)=A_\gamma^\dagger(t)A_\gamma(t)$ and $\hat
n_\gamma^{(b)}(t)=b_\gamma^\dagger(t)b_\gamma(t)=B_\gamma^\dagger(t)B_\gamma(t)$. Assuming that for $t=0$ both the populations are concentrated
in the cell 1\footnote{Of course this choice is not very relevant in the context of migration but is useful just to fix the ideas. The
application to migration will be considered for a larger lattice, where the situation is surely more realistic.}, $n_1^{(a)}(0)=n_1^{(b)}(0)=1$,
while $n_\alpha^{(a)}(0)=n_\alpha^{(b)}(0)=0$ for $\alpha=2,3,4$, we find the following results \be
\begin{aligned}
&n_1^{(a)}(t)=n_1^{(b)}(t)=\left(\cos(4t)\right)^4,\\
&n_2^{(a)}(t)=n_2^{(b)}(t)=\frac{1}{4}\left(\sin(8t)\right)^2,\\
&n_3^{(a)}(t)=n_3^{(b)}(t)=\frac{1}{4}\left(\sin(8t)\right)^2,\\
&n_4^{(a)}(t)=n_4^{(b)}(t)=\left(\sin(4t)\right)^4.\\
\end{aligned}
\label{36} \en These results look  reasonable because of the following considerations:
\begin{enumerate}
\item taking into account the fact that, because of our simplifying assumptions, all the parameters of $a$ and $b$ coincide, and since their initial conditions coincide as well, it is clear that the spreading of  the two populations must be identical;
\item at $t=0$ only the cell 1 is populated;
\item we observe that $n_2^{(a)}(t)=n_3^{(a)}(t)$ and $n_2^{(b)}(t)=n_3^{(b)}(t)$; this is not surprising since, because of the isotropy of $\R$, there is an equal probability for, say, a member of
$\Sc_1$  to move from cell 1 to cell 2 or to cell 3; so he can reach cell 4 only through cells 2 or 3, but not directly;
\item this explains why, if we plot $n_1^{(a)}(t)$, $n_2^{(a)}(t)$ and $n_4^{(a)}(t)$, we see that, for small values of $t$, $n_2^{(a)}(t)$ increases faster than $n_4^{(a)}(t)$ but, after some time, the opposite happens; this is because after cells 2 and 3 are populated, they both start contributing to the population of cell 4.
\end{enumerate}

Let us now move to $L=3$. In this case $\R$ is made up of 9 cells, and the only non--zero diffusion coefficients are (listing just one between
$p_{\alpha,\beta}$ and $p_{\beta,\alpha}$) $p_{1,2}$, $p_{1,4}$, $p_{2,5}$, $p_{2,3}$, $p_{3,6}$, $p_{4,5}$, $p_{4,7}$, $p_{5,6}$, $p_{5,8}$,
$p_{6,9}$, $p_{7,8}$ and $p_{8,9}$, which are all equal to 1. The 18 differential equations extending those in (\ref{35}) can be written as \be
\dot X_9=i \M_9 X_9,\qquad \M_9=2\tilde\gamma M_9-\lambda J_9, \label{37} \en where we have introduced the following vector and matrices:
$$
X_9=\left(
      \begin{array}{c}
        A_1 \\
        A_2 \\
        \dots \\
        \dots \\
        A_9 \\
        B_1 \\
        B_2 \\
        \dots \\
        \dots \\
        B_9 \\
      \end{array}
    \right),
    \quad N_9=\left(
                \begin{array}{ccccccccc}
                  0 & 1 & 0 & 1 & 0 & 0 & 0 & 0 & 0 \\
                  1 & 0 & 1 & 0 & 1 & 0 & 0 & 0 & 0 \\
                  0 & 1 & 0 & 0 & 0 & 1 & 0 & 0 & 0 \\
                  1 & 0 & 0 & 0 & 1 & 0 & 1 & 0 & 0 \\
                  0 & 1 & 0 & 1 & 0 & 1 & 0 & 1 & 0 \\
                  0 & 0 & 1 & 0 & 1 & 0 & 0 & 0 & 1 \\
                  0 & 0 & 0 & 1 & 0 & 0 & 0 & 1 & 0 \\
                  0 & 0 & 0 & 0 & 1 & 0 & 1 & 0 & 1 \\
                  0 & 0 & 0 & 0 & 0 & 1 & 0 & 1 & 0 \\
                \end{array}
              \right),
$$
$$
M_9=\left(
      \begin{array}{cc}
        N_9 & 0_9 \\
        0_9 & N_9 \\
      \end{array}
    \right),\qquad J_9=\left(
      \begin{array}{cc}
        0_9 & \1_9 \\
        \1_9 & 0_9 \\
      \end{array}
    \right),
$$
where $0_9$ and $\1_9$ are respectively the $9\times9$ null and identity matrices. Notice that $\M_9$ is a symmetric real matrix.

The generalization to larger $\R$ is straightforward. In this case we have \be \dot X_{L^2}=i \M_{L^2} X_{L^2},\qquad \M_{L^2}=2\tilde\gamma
M_{L^2}-\lambda J_{L^2}. \label{38} \en Here the transpose of $ X_{L^2}$ is $(A_1, A_2,\ldots, A_ {L^2}, B_1, B_2,\ldots, B_ {L^2})$, while
$0_{L^2}$, $\1_{L^2}$ and $J_{L^2}$ extend those above, and
$$M_{L^2}=\left(
\begin{array}{cc}
N_{L^2} & 0_{L^2} \\
0_{L^2} & N_{L^2} \\
\end{array}
\right).$$  Once again,  $\M_{L^2}$ is a symmetric real matrix. Of course, the explicit form of the matrix $N_{L^2}$ can be constructed extending the previous considerations: this matrix have all zero entries but those matrix elements corresponding to nearest neighbors, which assume as values 1.

The solution of equation (\ref{38}) is
\[
X_{L^2}(t)=\exp\left(i\,\M_{L^2}t\right)X_{L^2}(0).
\]
Let us call $d_{\alpha,\beta}(t)$ the generic entry of the matrix $\exp(i\,\M_{L^2}t)$, and let us assume that at $t=0$ the system is described
by the vector $\varphi_{\mathbf{n}^a,\mathbf{n}^b}$, where ${\bf n}^a=(n_1^a,n_2^a,\ldots,n_{L^2}^a)$ and  ${\bf
n}^b=(n_1^b,n_2^b,\ldots,n_{L^2}^b)$. Hence, the mean values of the time evolution of the number operators in the cell $\alpha$,
\[
\begin{aligned}
&N_\alpha^a(t)=\left<\varphi_{\mathbf{n}^a,\mathbf{n}^b},a_\alpha^\dagger(t) a_\alpha(t)\varphi_{\mathbf{n}^a,\mathbf{n}^b}\right>=\left<\varphi_{\mathbf{n}^a,\mathbf{n}^b},A_\alpha^\dagger(t) A_\alpha(t)\varphi_{\mathbf{n}^a,\mathbf{n}^b}\right>,\\
&N_\alpha^b(t)=\left<\varphi_{\mathbf{n}^a,\mathbf{n}^b},b_\alpha^\dagger(t) b_\alpha(t)\varphi_{\mathbf{n}^a,\mathbf{n}^b}\right>=\left<\varphi_{\mathbf{n}^a,\mathbf{n}^b},B_\alpha^\dagger(t) B_\alpha(t)\varphi_{\mathbf{n}^a,\mathbf{n}^b}\right>,
\end{aligned}
\]
can be written as \be \label{39}
\begin{aligned}
&N_\alpha^a(t)=\sum_{\theta=1}^{L^2}|d_{\alpha,\theta}(t)|^2\,n_\theta^a+
\sum_{\theta=1}^{L^2}|d_{\alpha,L^2+\theta}(t)|^2\,n_\theta^b,\\
&N_\alpha^b(t)=\sum_{\theta=1}^{L^2}|d_{L^2+\alpha,\theta}(t)|^2\,n_\theta^a+
\sum_{\theta=1}^{L^2}|d_{L^2+\alpha,L^2+\theta}(t)|^2\,n_\theta^b.
\end{aligned}
\en

\begin{figure}
\begin{center}
\includegraphics[width=0.40\textwidth]{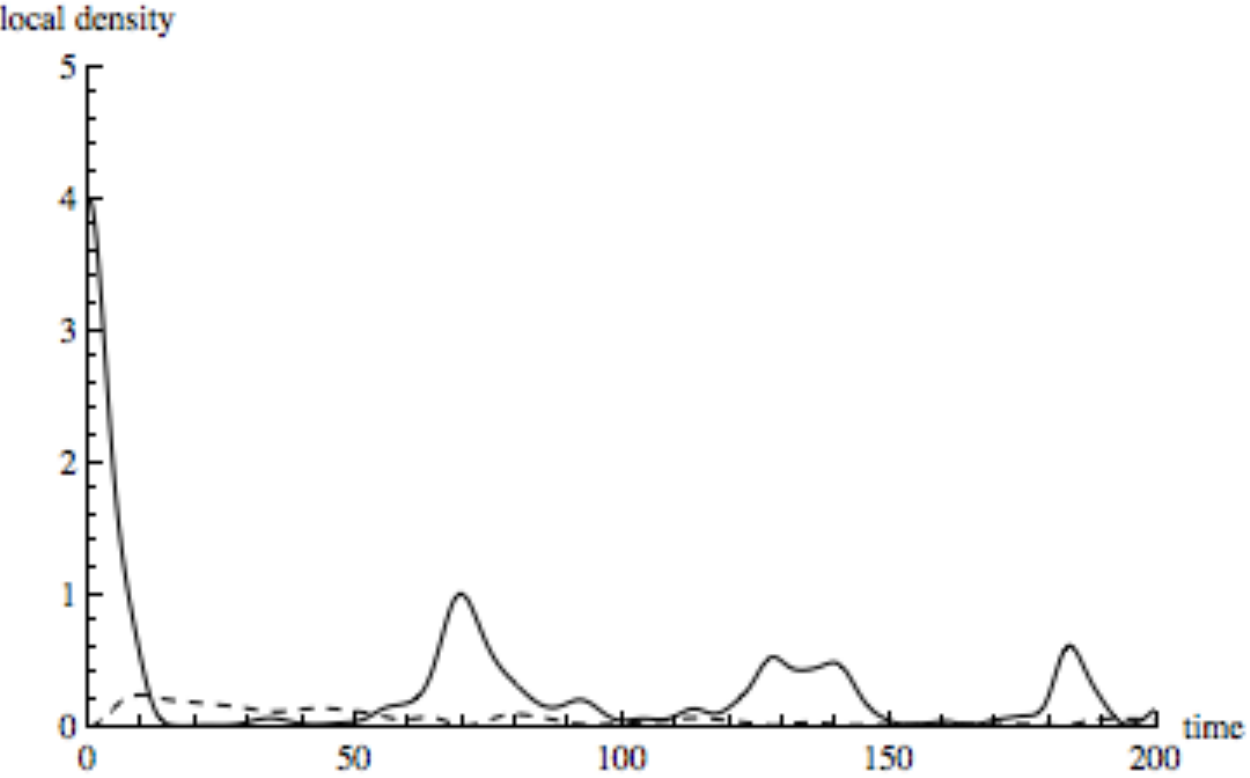}\;\includegraphics[width=0.40\textwidth]{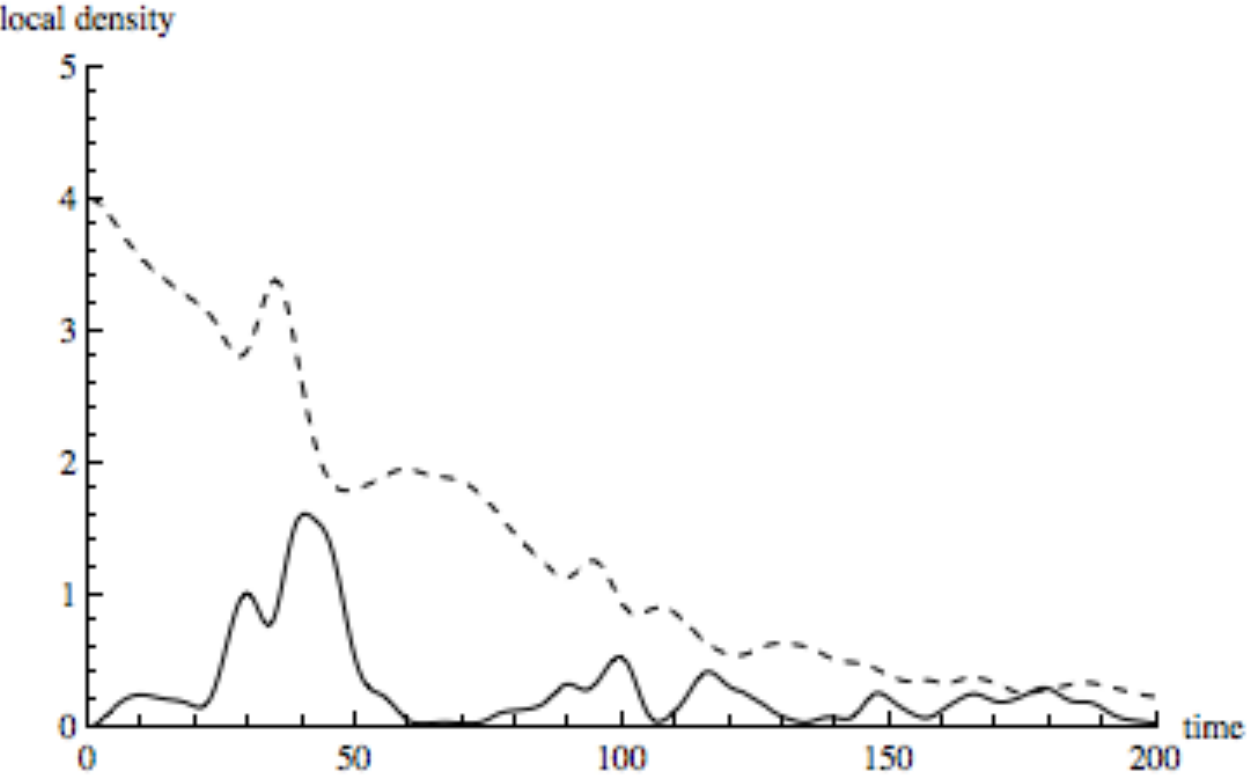}\\
\includegraphics[width=0.40\textwidth]{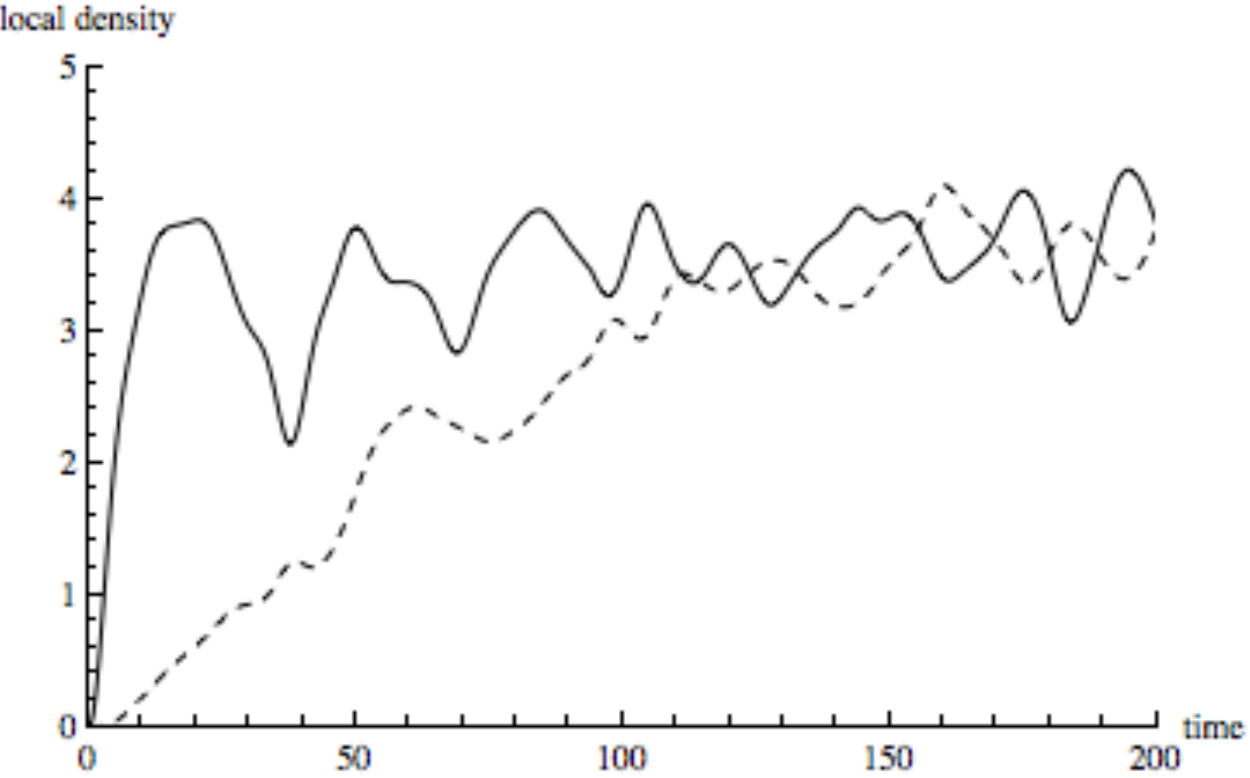}
\end{center}
\caption{\label{figA1}\footnotesize Evolution of local densities (solid line for $\Sc_1$ and dashed line for $\Sc_2$). Africa: top--left (a); Europe: top--right (b); Mediterranean Sea: bottom (c). $\gamma_a=0.1$, $\gamma_b=0.004$, $\omega^a_\alpha=1$, $\omega^b_\alpha=0.3$, $\lambda_\alpha=0.05$, $\forall \alpha\in\R$.}
\end{figure}

\begin{figure}
\begin{center}
\includegraphics[width=0.40\textwidth]{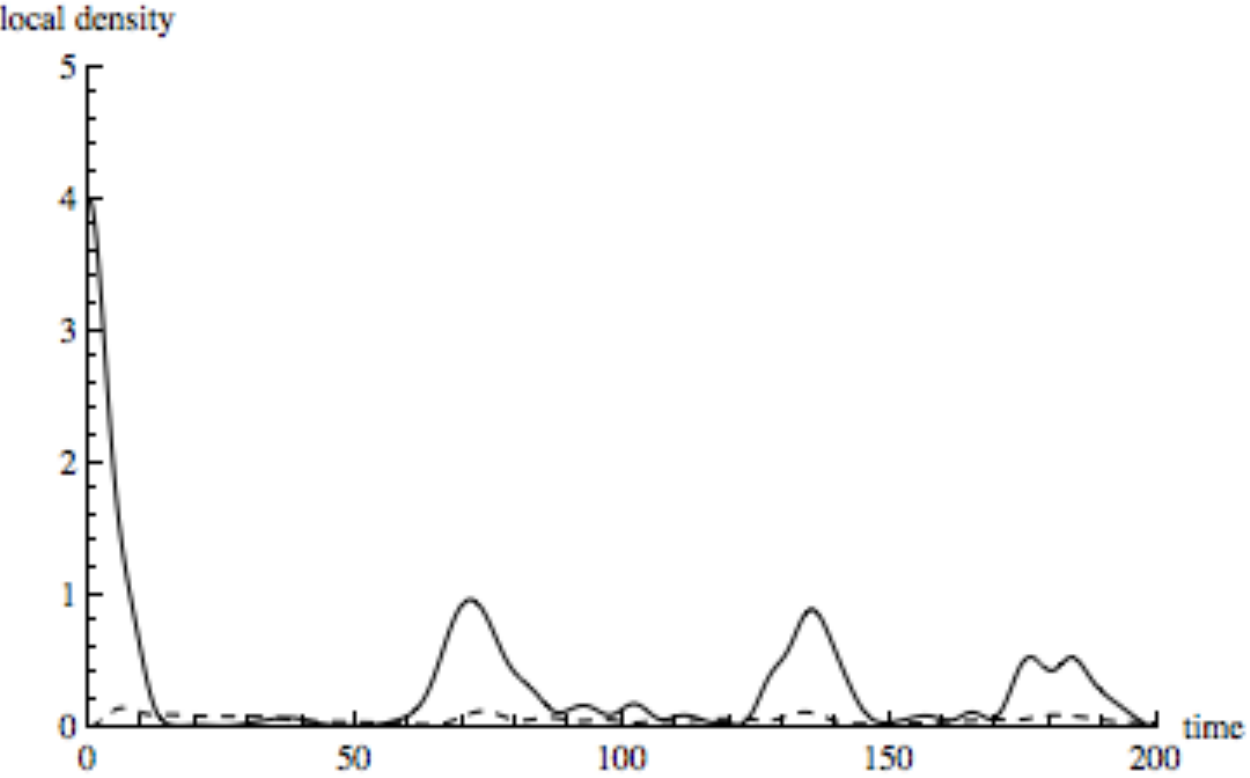}\;\includegraphics[width=0.40\textwidth]{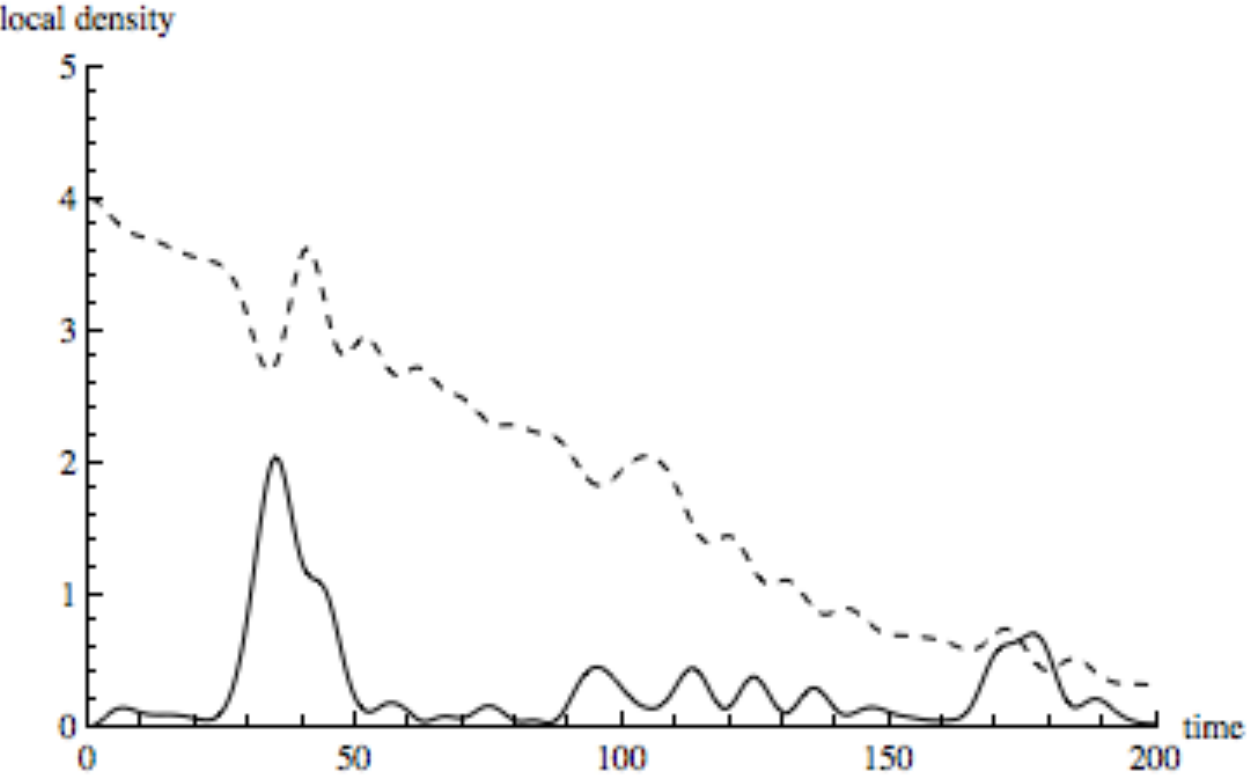}\\
\ \\
\includegraphics[width=0.40\textwidth]{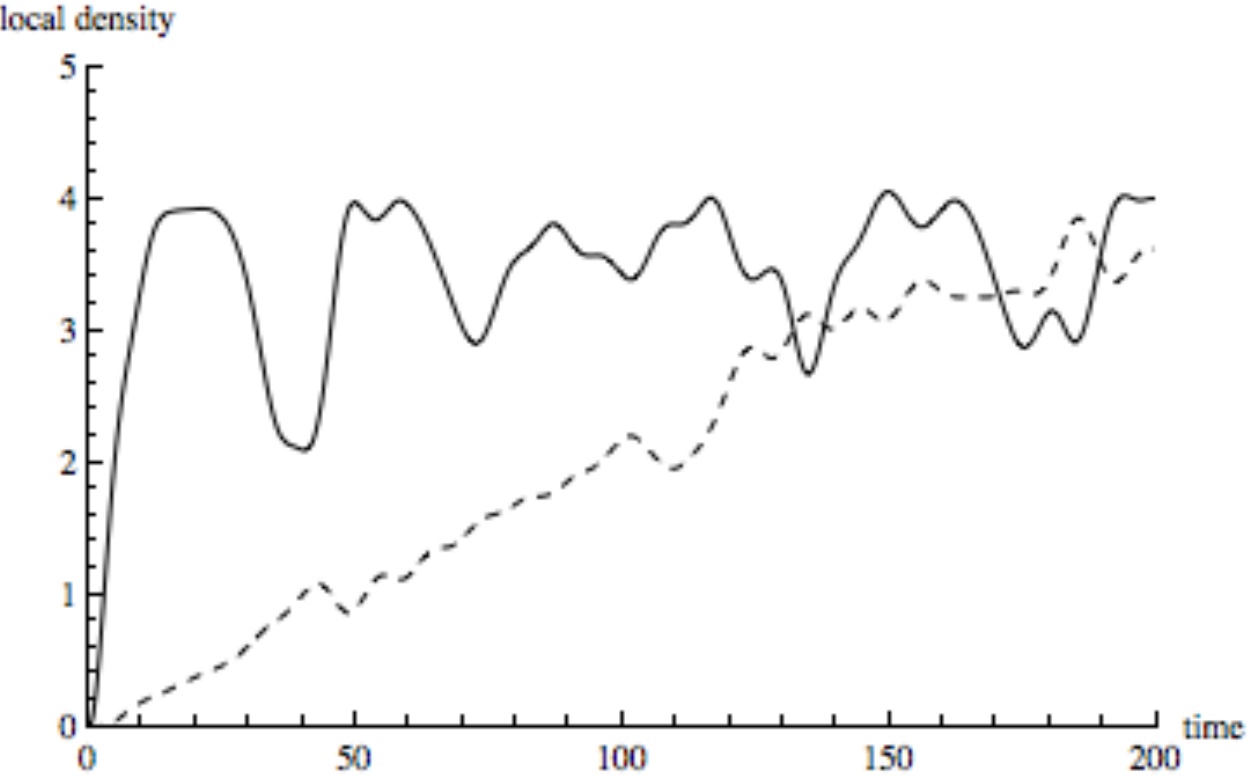}
\end{center}
\caption{\label{figA2} \footnotesize Evolution of local densities (solid line for $\Sc_1$ and dashed line for $\Sc_2$). Africa: top--left (a); Europe: top--right (b); Mediterranean Sea: bottom (c). $\gamma_a=0.1$, $\gamma_b=0.004$, $\omega^a_\alpha=1$, $\omega^b_\alpha=1$, $\lambda_\alpha=0.05$, $\forall \alpha\in\R$.}
\end{figure}

\begin{figure}
\begin{center}
\includegraphics[width=0.40\textwidth]{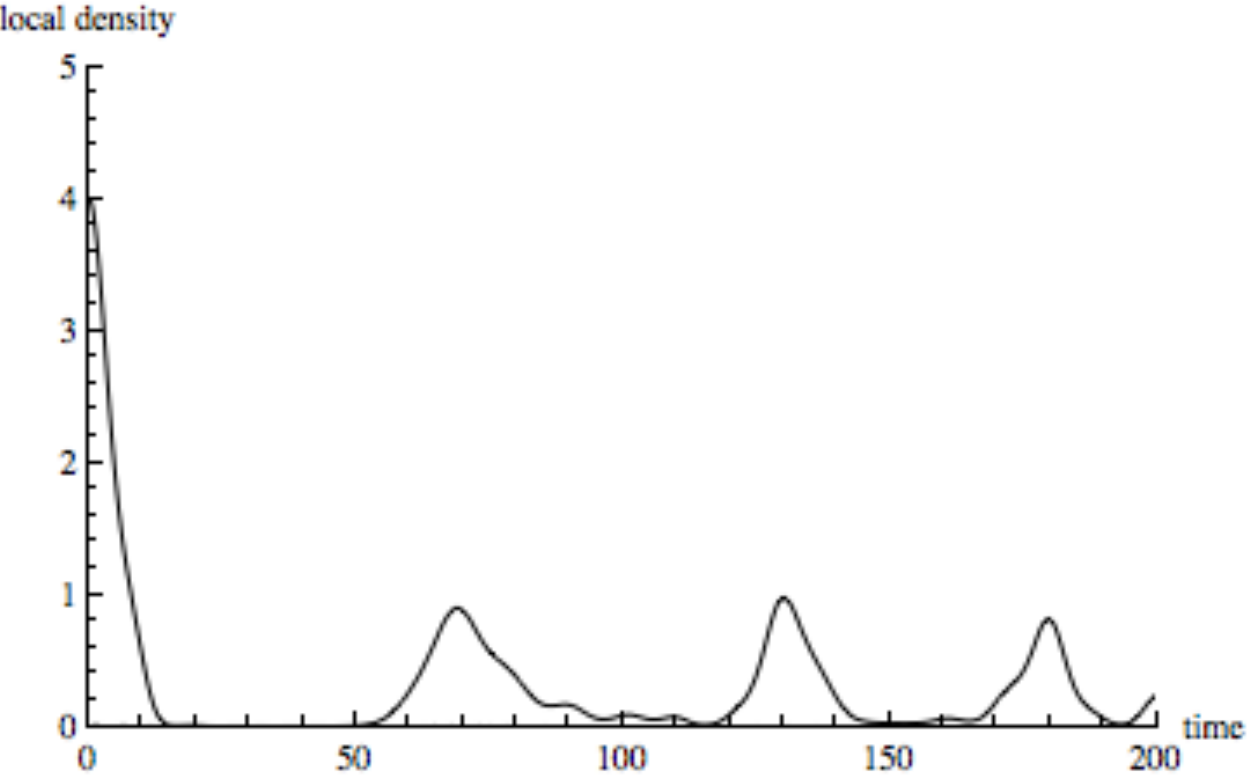}\;\includegraphics[width=0.40\textwidth]{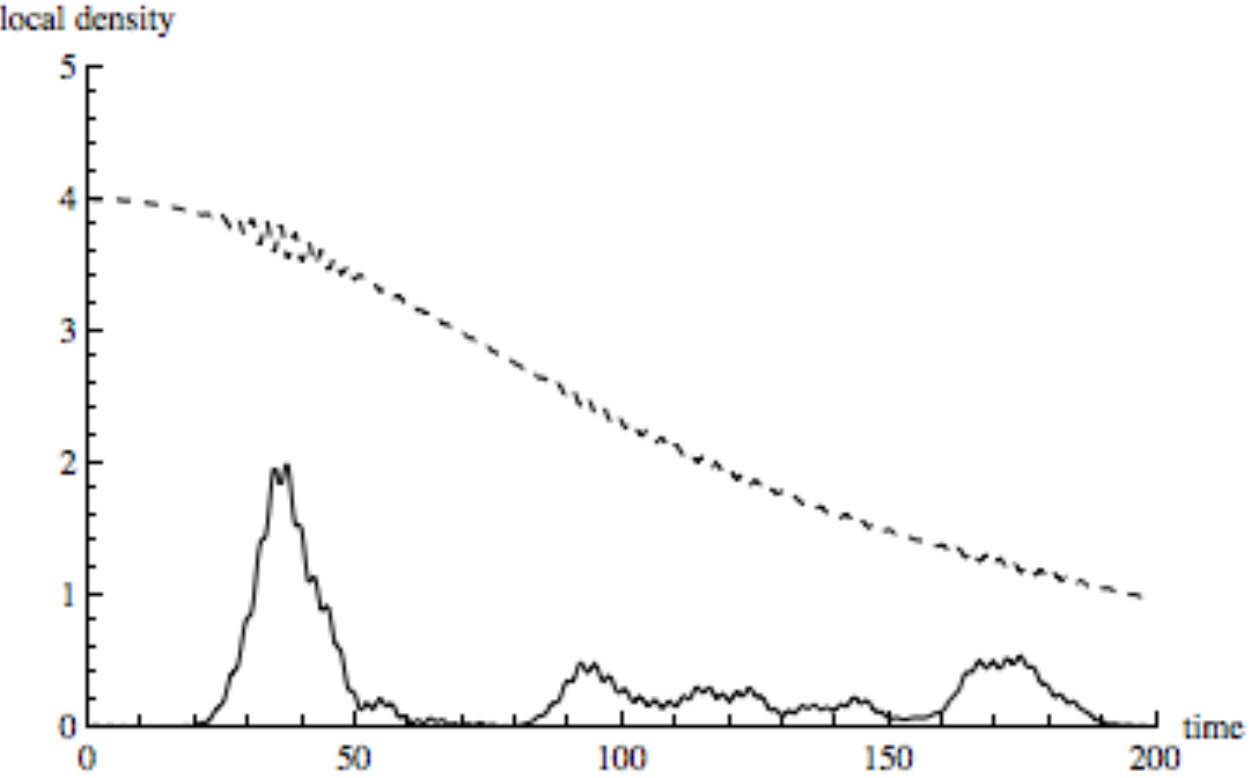}\\
\ \\
\includegraphics[width=0.40\textwidth]{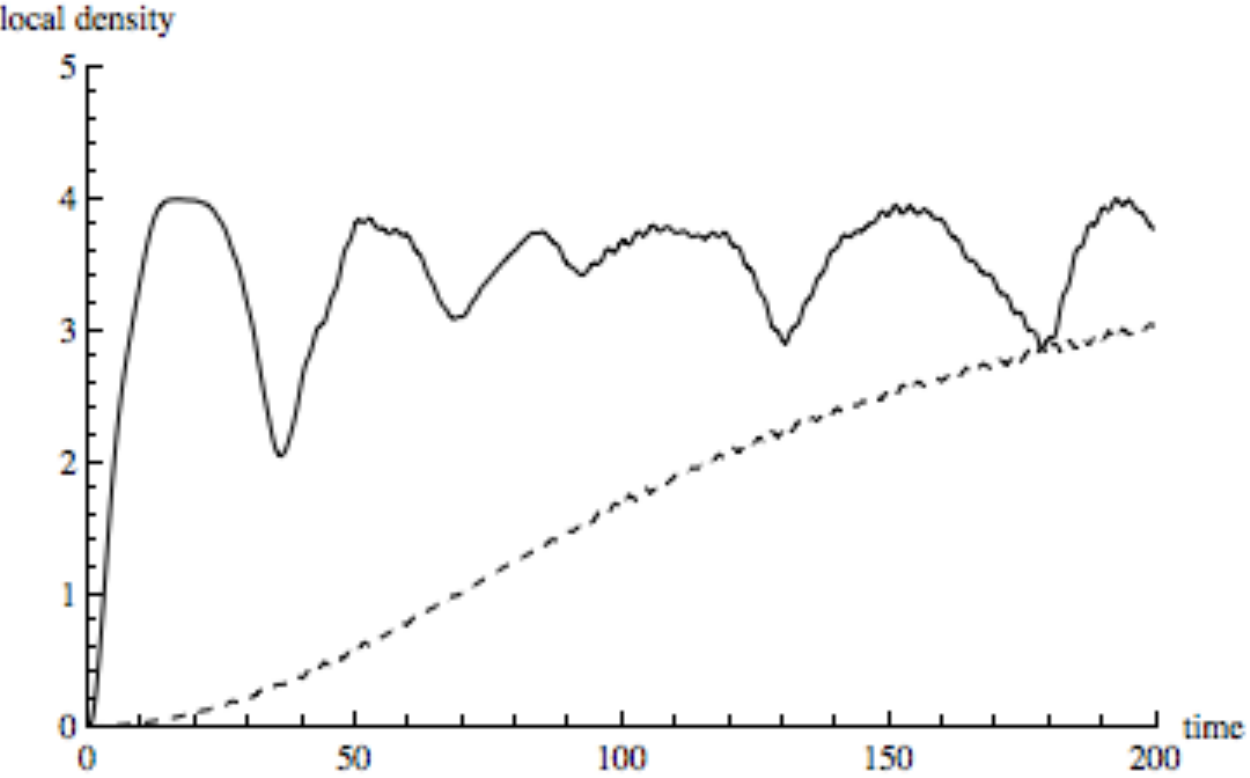}
\end{center}
\caption{\label{figA3} \footnotesize Evolution of local densities (solid line for $\Sc_1$ and dashed line for $\Sc_2$). Africa: top--left (a); Europe: top--right (b); Mediterranean Sea: bottom (c). $\gamma_a=0.1$, $\gamma_b=0.004$, $\omega^a_\alpha=1$, $\omega^b_\alpha=3$, $\lambda_\alpha=0.05$, $\forall \alpha\in\R$.}
\end{figure}

\begin{figure}
\begin{center}
\includegraphics[width=0.40\textwidth]{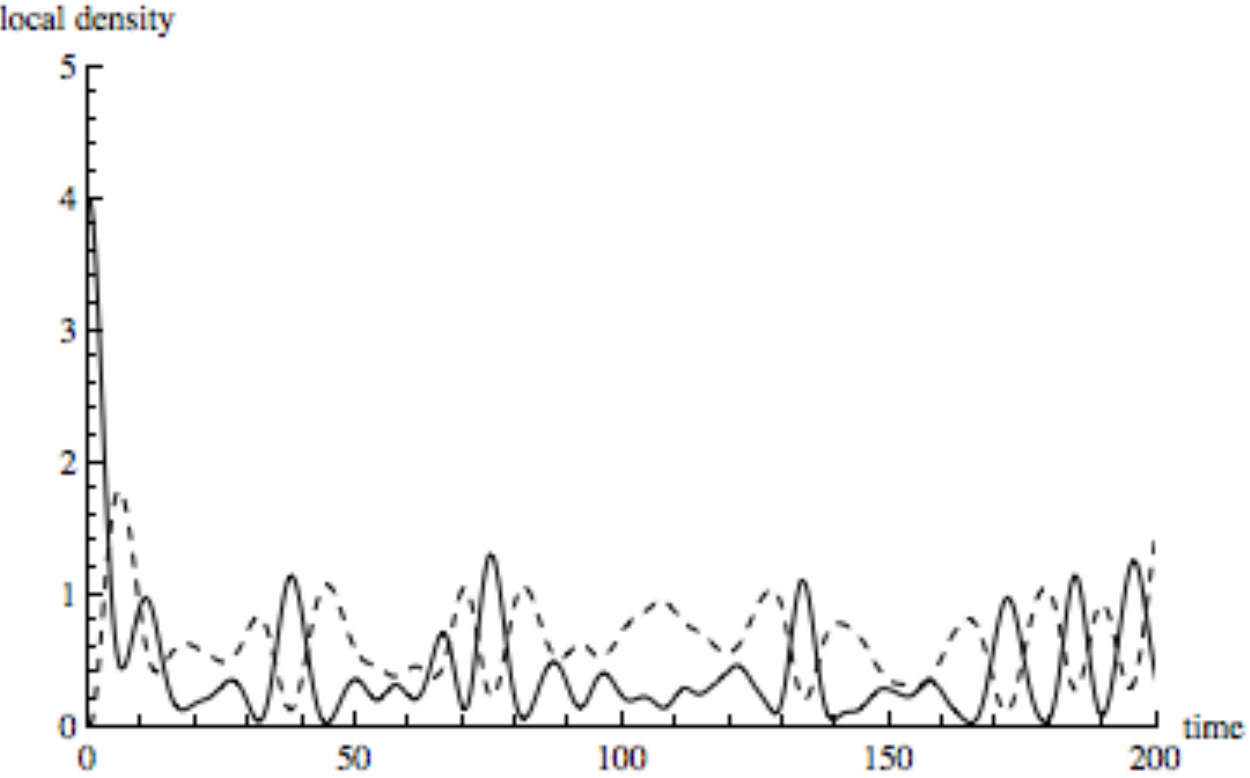}\;\includegraphics[width=0.40\textwidth]{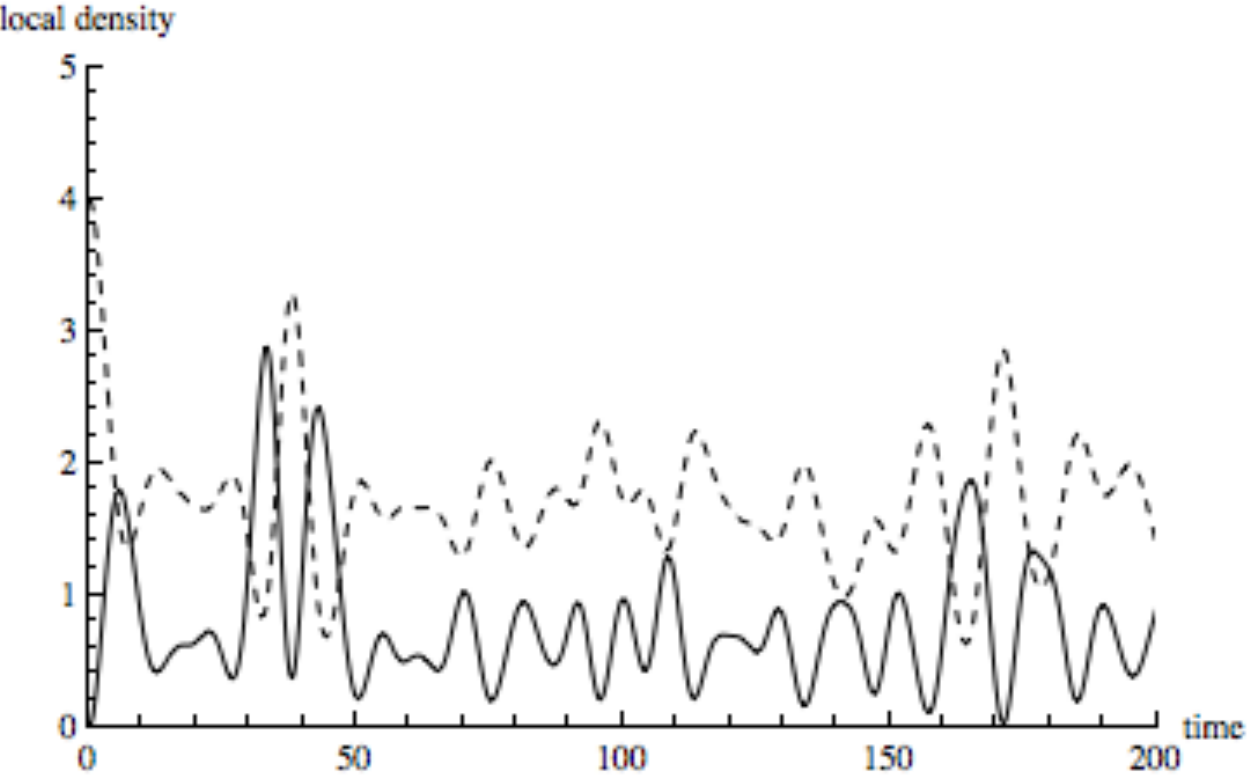}\\
\ \\
\includegraphics[width=0.40\textwidth]{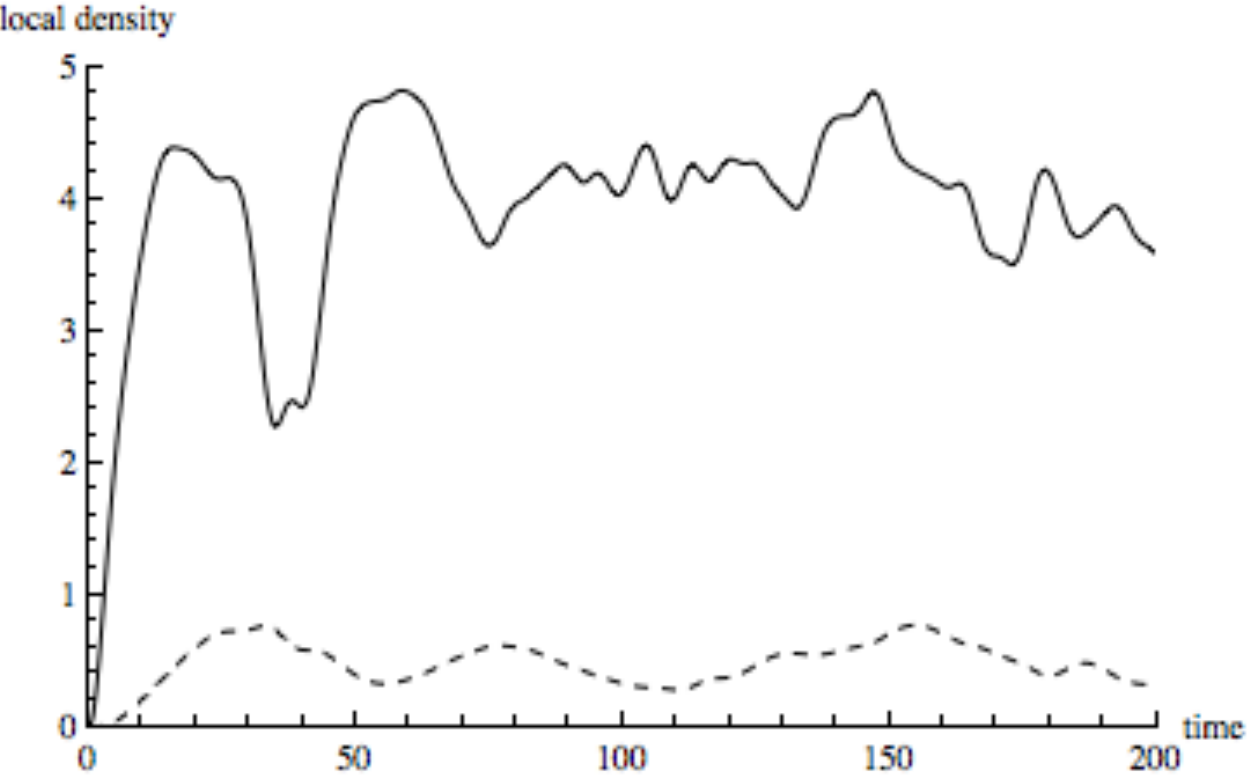}
\end{center}
\caption{\label{figA4} \footnotesize Evolution of local densities (solid line for $\Sc_1$ and dashed line for $\Sc_2$). Africa: top--left (a); Europe: top--right (b); Mediterranean Sea: bottom (c). $\gamma_a=0.1$, $\gamma_b=0.004$, $\omega^a_\alpha=1$, $\omega^b_\alpha=0.3$, $\forall \alpha\in\R$; $\lambda_\alpha=0.2$ for $\alpha\in\R_1\cup\R_2$ and $\lambda_\alpha=0.05$ for $\alpha\in\R_3$.}
\end{figure}

\begin{figure}
\begin{center}
\includegraphics[width=0.40\textwidth]{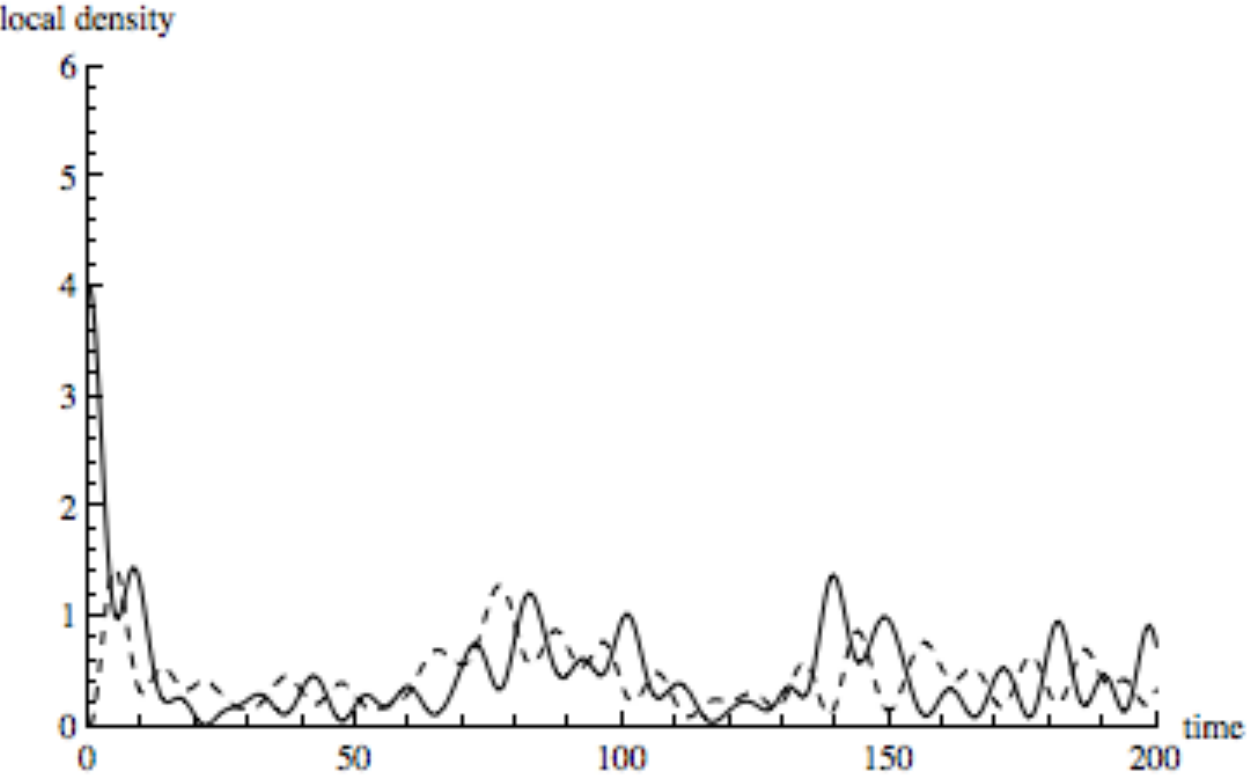}\;\includegraphics[width=0.40\textwidth]{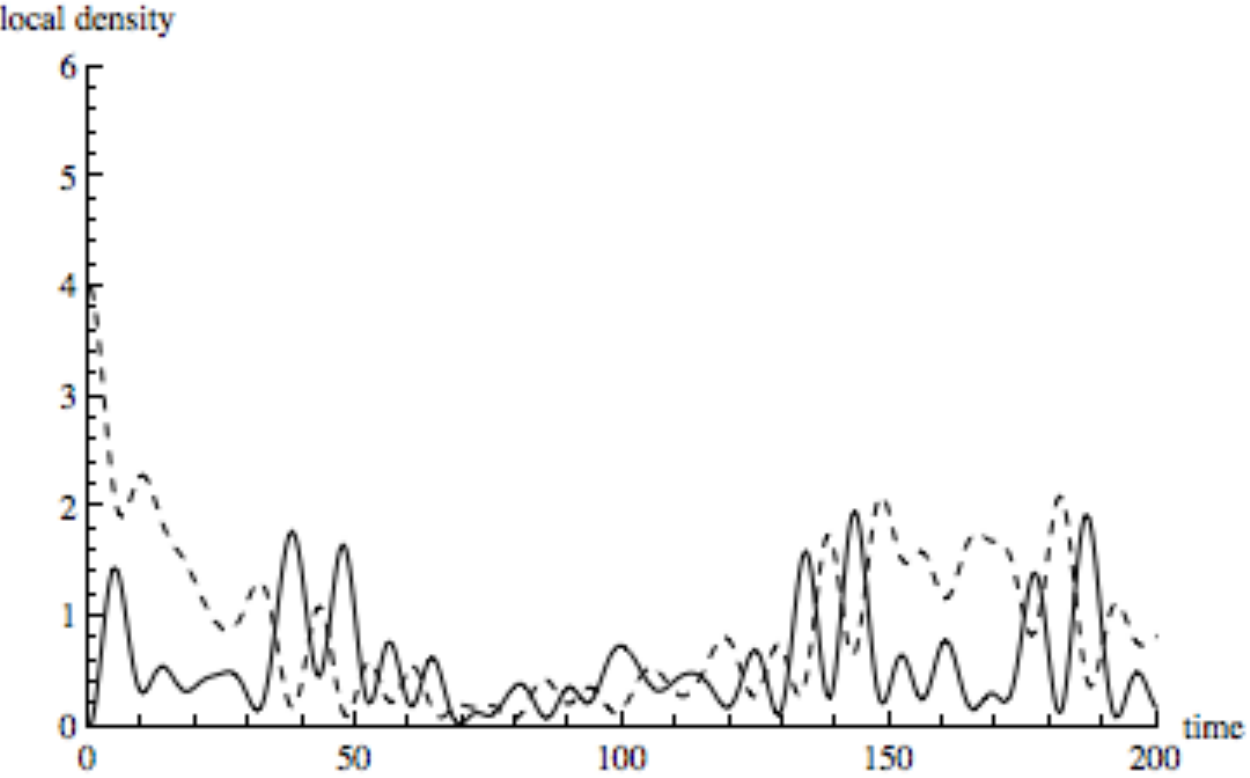}\\
\ \\
\includegraphics[width=0.40\textwidth]{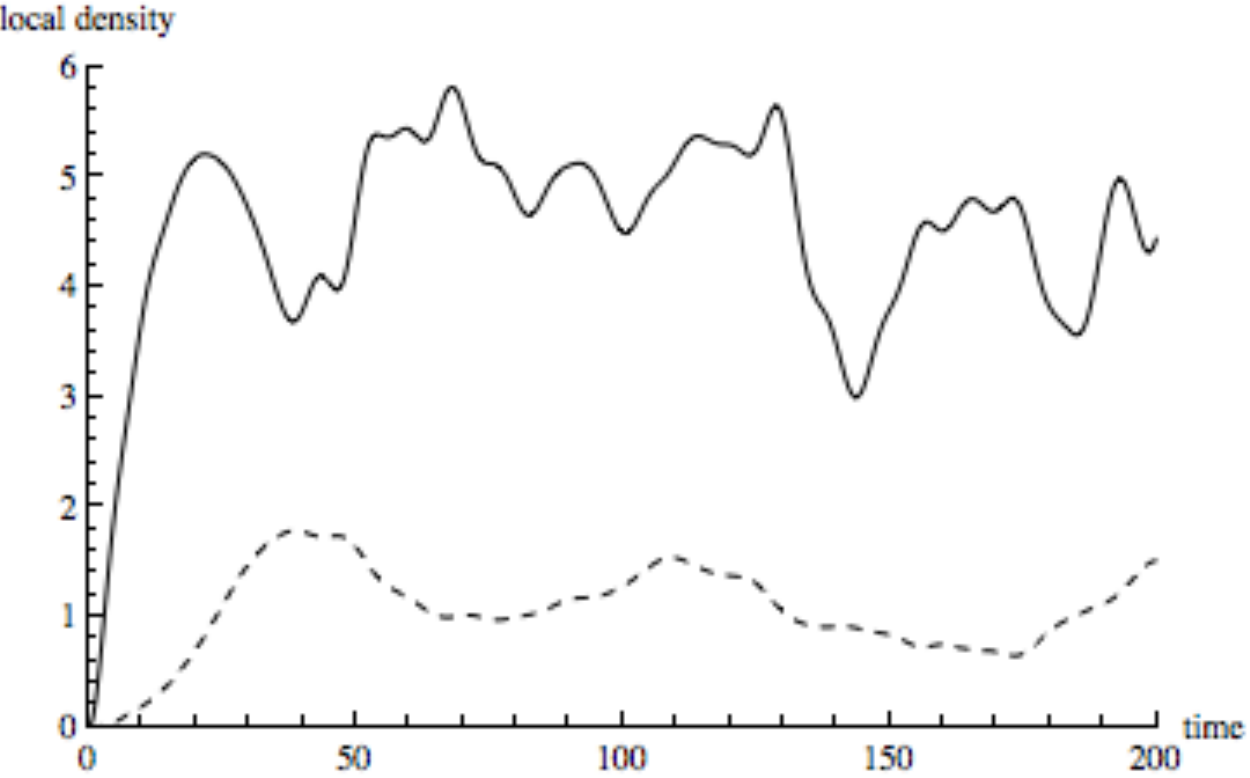}
\end{center}
\caption{\label{figA5} \footnotesize Evolution of local densities (solid line for $\Sc_1$ and dashed line for $\Sc_2$). Africa: top--left (a); Europe: top--right (b); Mediterranean Sea: bottom (c). $\gamma_a=0.1$, $\gamma_b=0.004$, $\omega^a_\alpha=1$, $\omega^b_\alpha=1$ $\forall \alpha\in\R$; $\lambda_\alpha=0.2$ for $\alpha\in\R_1\cup\R_2$ and $\lambda_\alpha=0.05$ for $\alpha\in\R_3$.}
\end{figure}

\begin{figure}
\begin{center}
\includegraphics[width=0.40\textwidth]{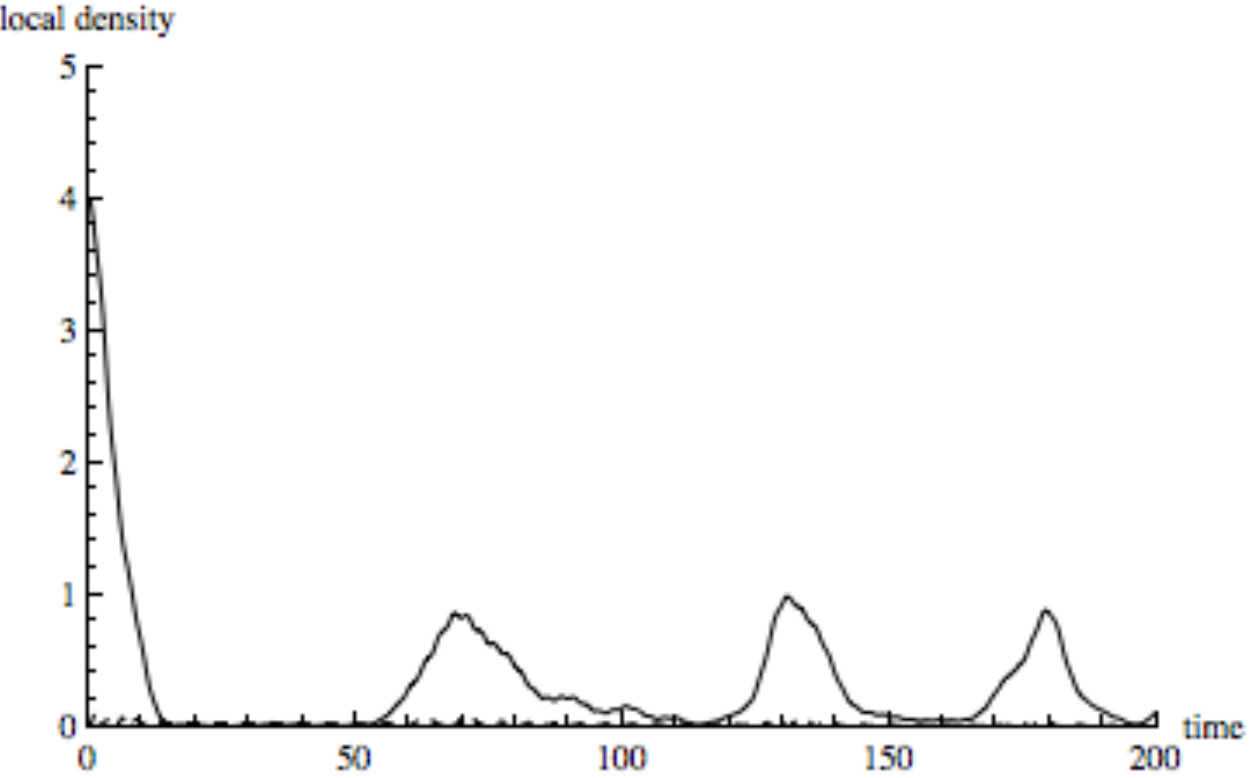}\;\includegraphics[width=0.40\textwidth]{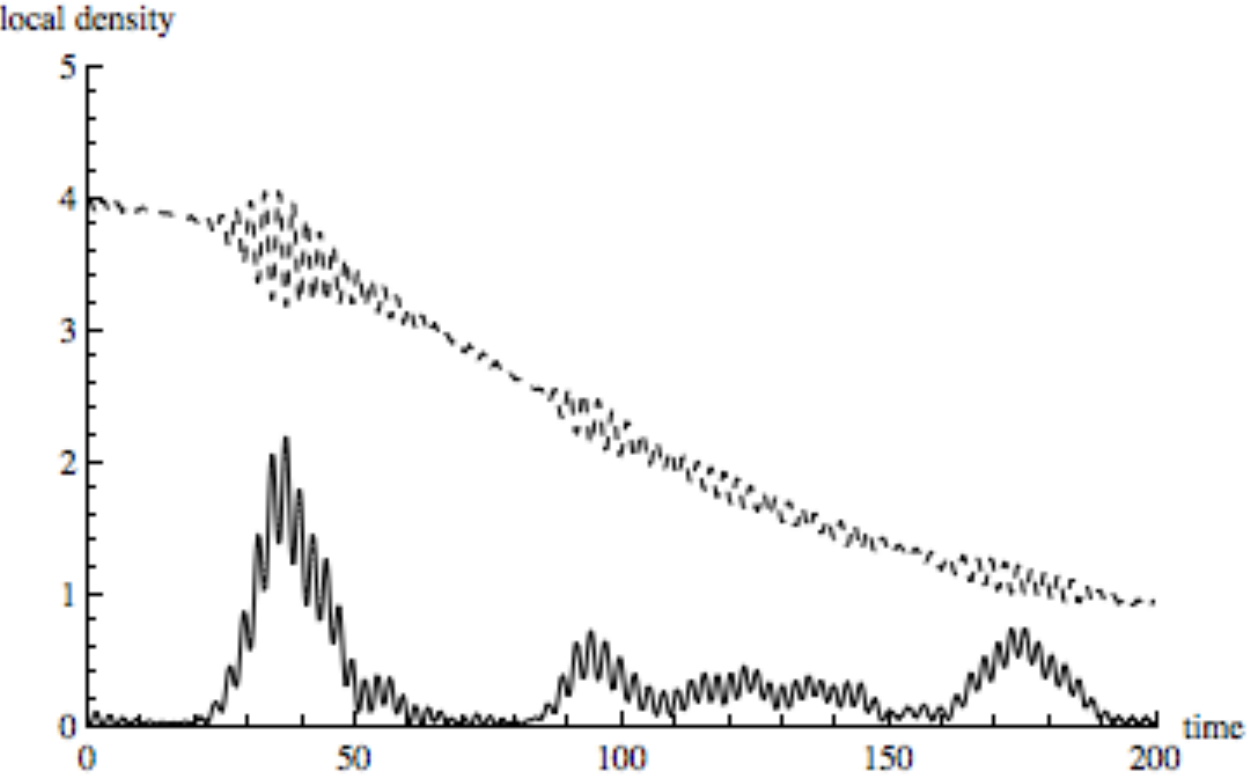}\\
\ \\
\includegraphics[width=0.40\textwidth]{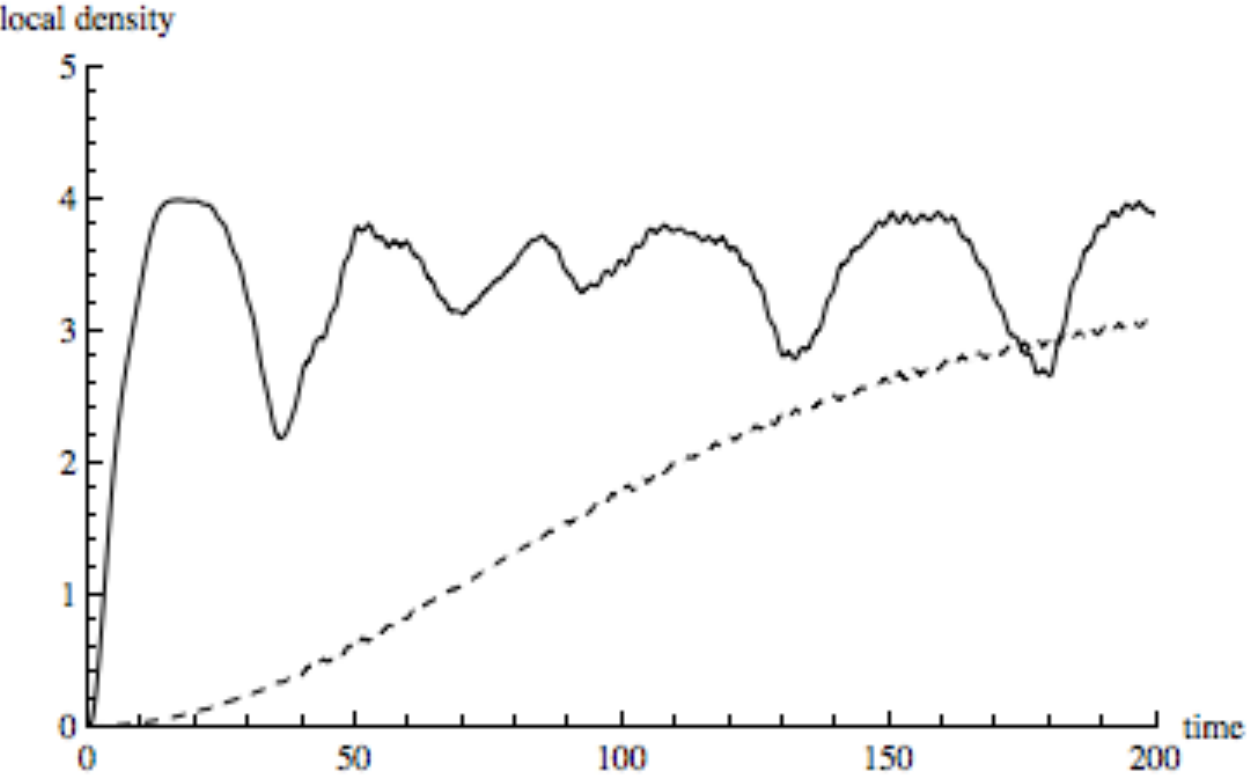}
\end{center}
\caption{\label{figA6} \footnotesize Evolution of local densities. Africa (solid line for $\Sc_1$ and dashed line for $\Sc_2$). Africa: top--left (a); Europe: top--right (b); Mediterranean Sea: bottom (c). $\gamma_a=0.1$, $\gamma_b=0.004$, $\omega^a_\alpha=1$, $\omega^b_\alpha=2$ $\forall \alpha\in\R$; $\lambda_\alpha=0.2$ for $\alpha\in\R_1\cup\R_2$ and $\lambda_\alpha=0.05$ for $\alpha\in\R_3$.}
\end{figure}

\begin{figure}
\begin{center}
\includegraphics[width=0.40\textwidth]{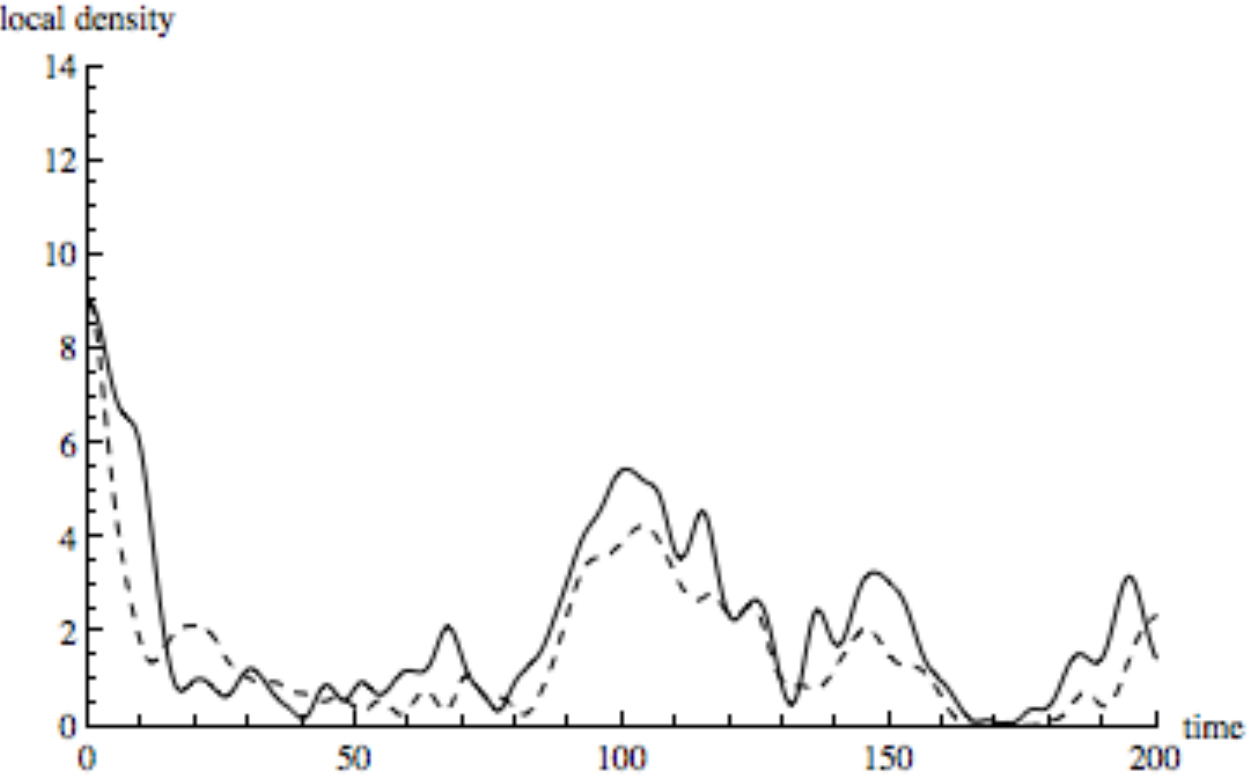}\;\includegraphics[width=0.40\textwidth]{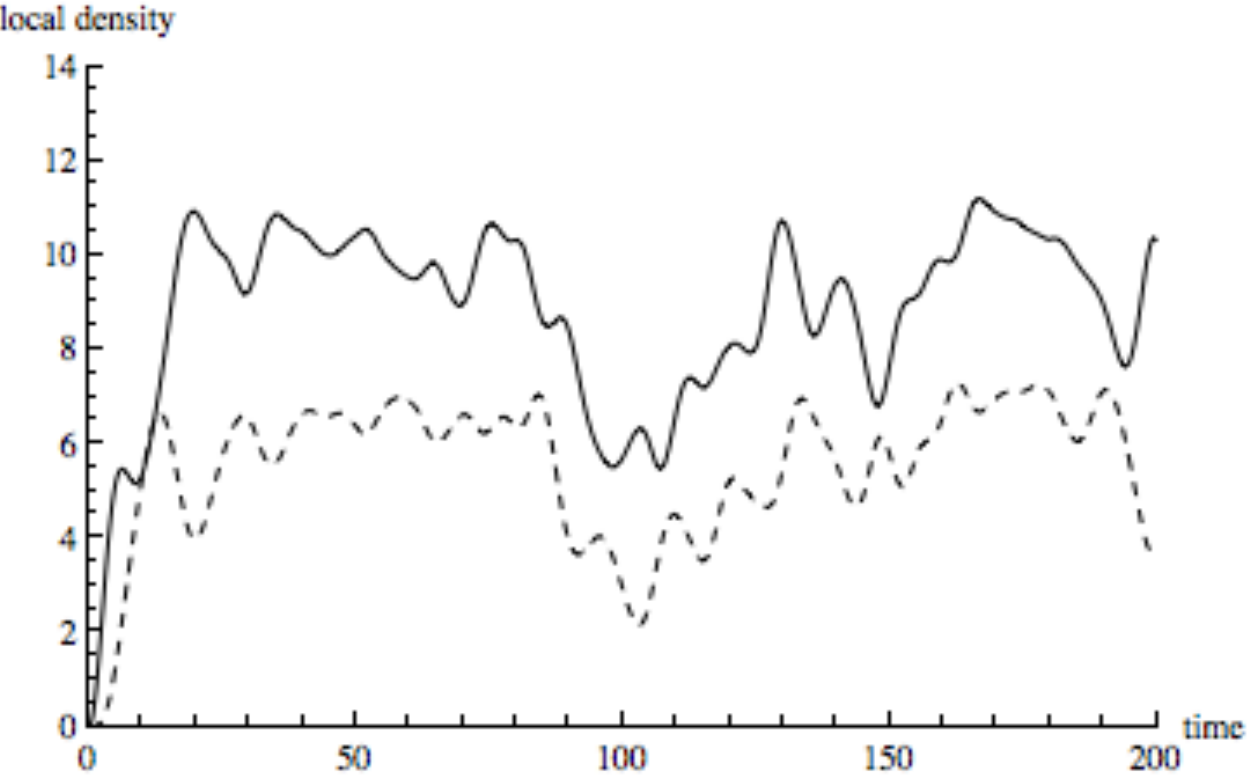}\\
\includegraphics[width=0.40\textwidth]{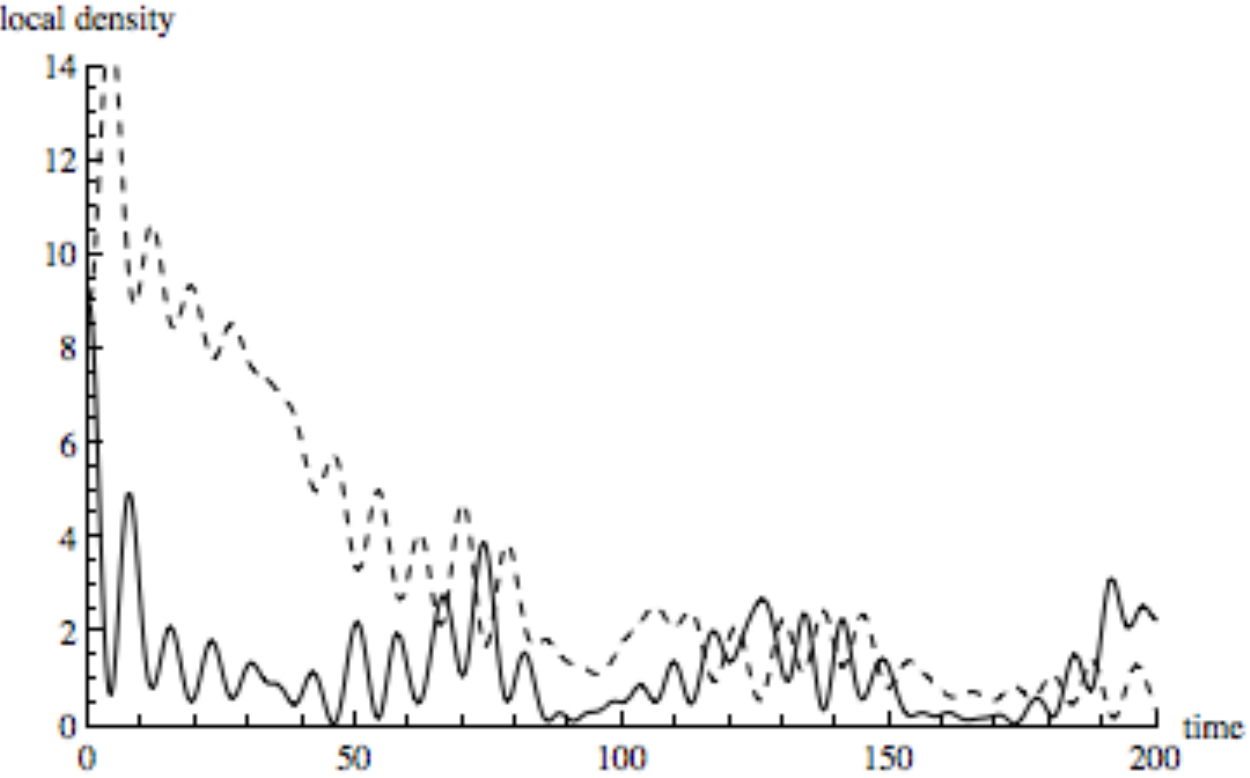}\;\includegraphics[width=0.40\textwidth]{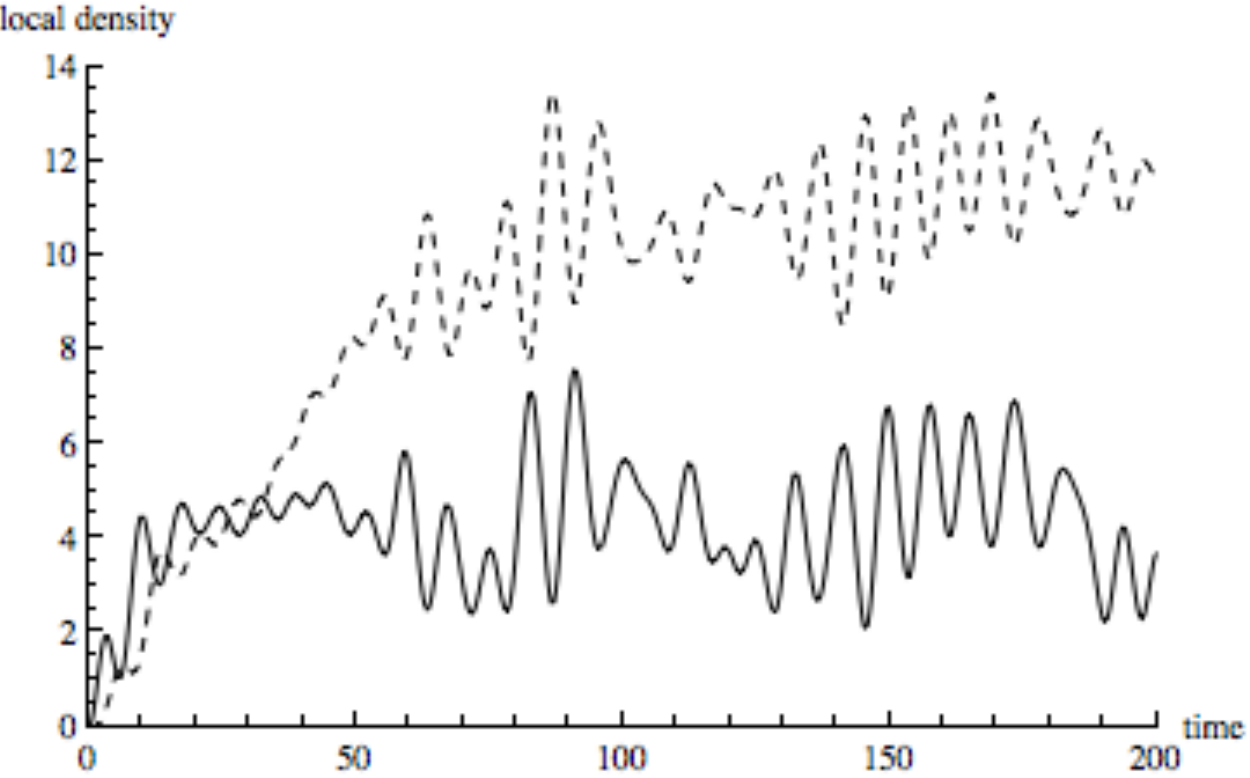}\\
\includegraphics[width=0.40\textwidth]{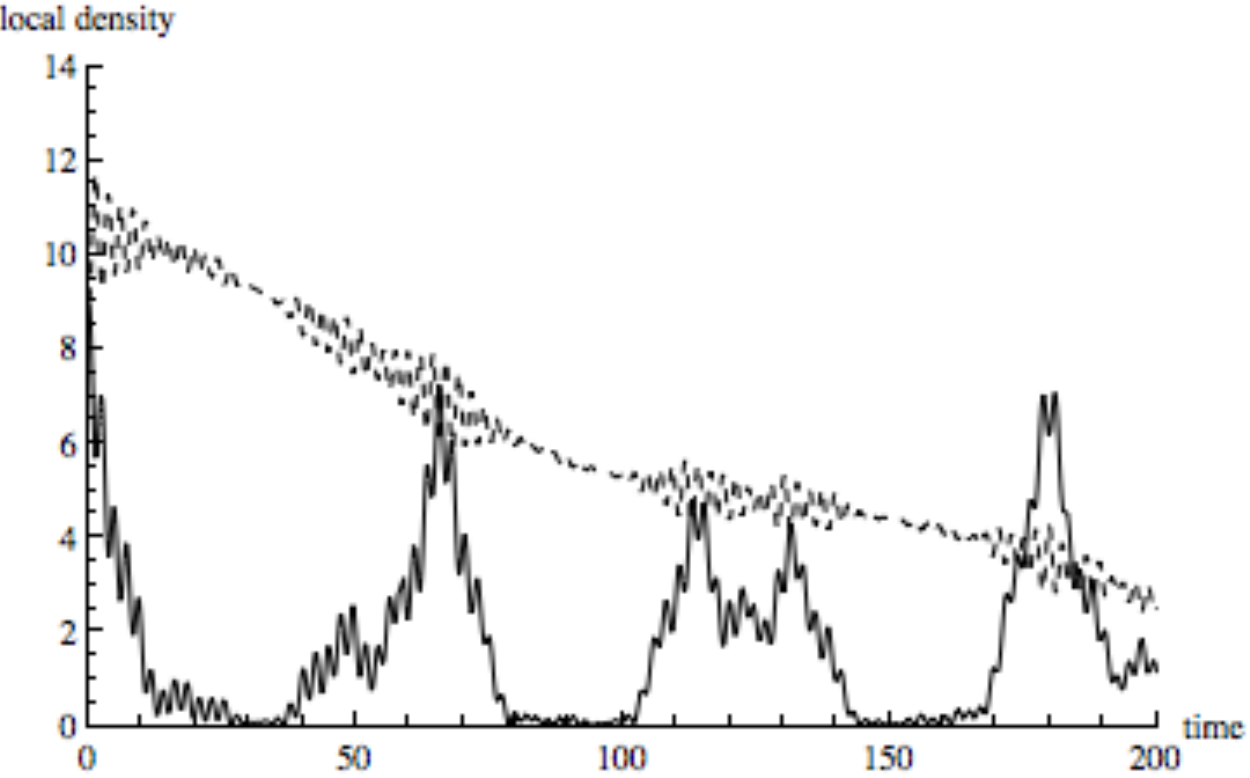}\;\includegraphics[width=0.40\textwidth]{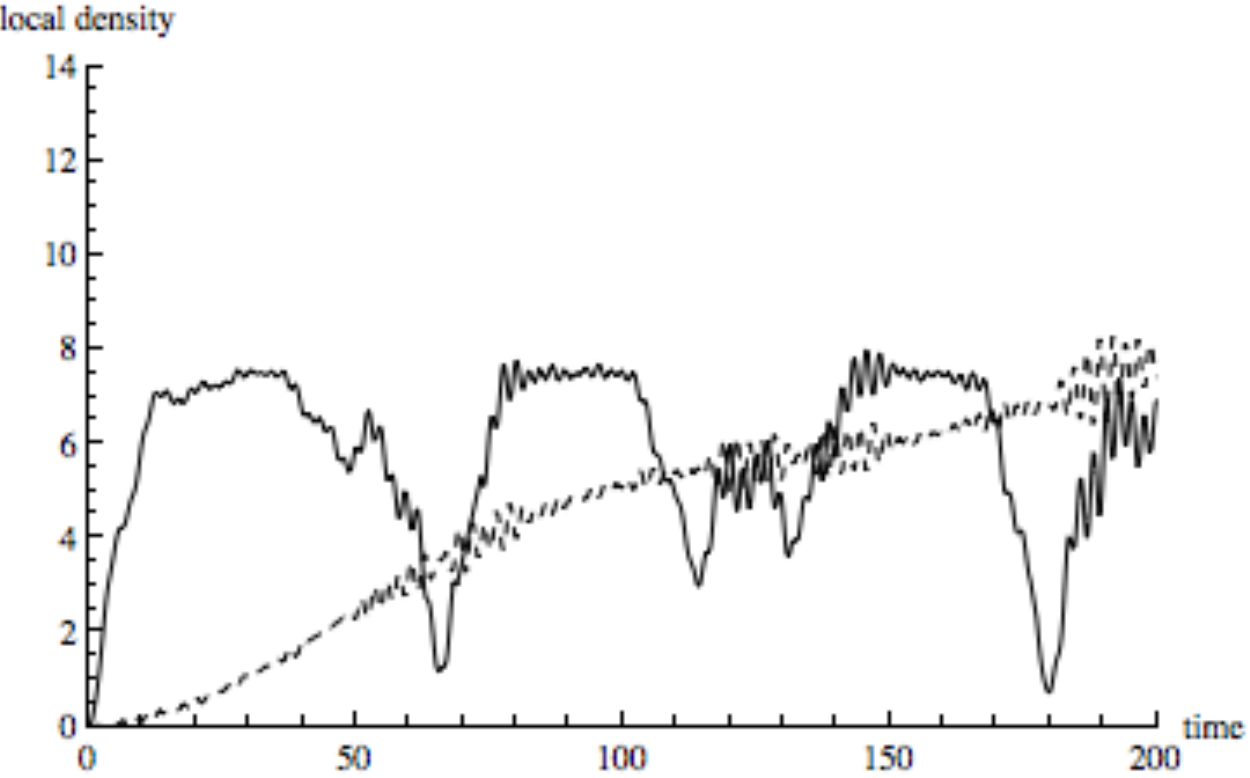}\\
\end{center}
\caption{\label{figA7} \footnotesize Evolution of local densities  (solid line for $\Sc_1$ and dashed line for $\Sc_2$). Inside $\R_c$ (left), Outside $\R_c$ (right). $\gamma_a=0.1$, $\gamma_b=0.004$, $\lambda_\alpha=0.2$ for $\alpha\in\R$. First row: $\omega^a_\alpha=1$, $\omega^b_\alpha=0.3$, $\forall \alpha\in\R$. Second row: $\omega^a_\alpha=1$, $\omega^b_\alpha=1$, $\forall \alpha\in\R$. Third row: $\omega^a_\alpha=1$, $\omega^b_\alpha=3$, $\forall \alpha\in\R$.}
\end{figure}

\subsection{Back to the general case: migration}

The same strategy which produces solution (\ref{39}) can be used to solve system (\ref{34}). In this case, (\ref{38}) is replaced by a similar
equation, \be \dot X_{L^2}=i \K_{L^2} X_{L^2}, \label{310}\en where $\K_{L^2}=2T_{L^2}-P_{L^2}$, with $T_{L^2}$ and $P_{L^2}$ two $L^2\times
L^2$ matrices defined as follows:
\[
T_{L^2}=\left(
          \begin{array}{cc}
            N_{L^2}\mbox{ with 1 replaced by }\gamma_a & 0 \\
            0 & N_{L^2}\mbox{ with 1 replaced by }\gamma_b \\
          \end{array}
        \right)
\]
and
\[
P_{L^2}=\left(
          \begin{array}{cc}
            \Omega^{(a)} & \Lambda \\
            \Lambda & \Omega^{(b)} \\
          \end{array}
        \right).
\]
Here we have introduced the following matrices: $\Omega^{(a)}=\hbox{diag}\{\omega_1^a,\omega_2^a,\ldots,\omega_{L^2}^a\}$,
$\Omega^{(b)}=\hbox{diag}\{\omega_1^b,\omega_2^b,\ldots,\omega_{L^2}^b\}$, and
$\Lambda=\hbox{diag}\{\lambda_1,\lambda_2,\ldots,\lambda_{L^2}\}$.

The solution of equation (\ref{310}) is
\[
X_{L^2}(t)=\exp\left(i\,\K_{L^2}t\right)X_{L^2}(0).
\]
Calling $f_{\alpha,\beta}(t)$ the generic entry of the matrix $\exp\left(i\,\K_{L^2}t\right)$, and repeating the same procedure as above, we
get \be \label{311}
\begin{aligned}
&N_\alpha^a(t)=\sum_{\theta=1}^{L^2}|f_{\alpha,\theta}(t)|^2\,n_\theta^a+
\sum_{\theta=1}^{L^2}|f_{\alpha,L^2+\theta}(t)|^2\,n_\theta^b,\\
&N_\alpha^b(t)=\sum_{\theta=1}^{L^2}|f_{L^2+\alpha,\theta}(t)|^2\,n_\theta^a+
\sum_{\theta=1}^{L^2}|f_{L^2+\alpha,L^2+\theta}(t)|^2\,n_\theta^b.
\end{aligned}
\en

These formulas are used to deduce the local densities of the two populations $\Sc_1$ and $\Sc_2$ in three different regions. The first one,
$\R_{1}$, corresponding to cells 1, 2, $L+1$ and $L+2$ (bottom--left corner of $\R$), is that part of $\R$ where all the members of $\Sc_1$ are
originally ({\em i.e.}, at $t=0$) localized. Population $\Sc_2$, at $t=0$, is assumed to be localized in the four cells $L^2-L-1$, $L^2-L$,
$L^2-1$ and $L^2$, the region $\R_2$ (top--right corner of $\R$). All the other cells form $\R_3$, that part of $\R$ which must be crossed by
the populations to reach the other region of the lattice. Just to fix the ideas, we could think of $\Sc_1$ and $\Sc_2$ as people from Africa
($\R_1$) and Europe ($\R_2$), respectively, and the Mediterranean sea as the region $\R_3$. We also fix $L=11$. In Figures
\ref{figA1}--\ref{figA6} we plot the two {\em local densities} (the sum of the densities in the different cells) for $\Sc_1$ (solid line) and
$\Sc_2$ (dashed line), in $\R_1$ (top--left plot), $\R_2$ (top--right plot) and $\R_3$ (bottom plot), for different choices of the parameters
and for the same initial conditions given above.

\begin{rem}
The reason why we are talking here of Africa and Europe is very much related to what we have experienced in Italy, and in
Sicily in particular, during this last year, with all the people moving from Africa, and from Libya in particular, and reaching Lampedusa
first, and Europe soon after. In the past century, to a similar migration process took part a lot of people coming from Sicily (among whom many
of our relatives) and moving to America looking for a better life. Many of them reached a reasonable well-being, and some of them returned back
to their own villages. This is, by the way, essentially what our results show.
\end{rem}


In particular, in Figures \ref{figA1}-\ref{figA3}  the parameter $\lambda_\alpha$ is taken to be equal, $\lambda_\alpha=0.05$, in all the cells
of $\R$. On the other hand, in Figures \ref{figA4}-\ref{figA6},  $\lambda_\alpha=0.05$ in $\R_3$ while $\lambda_\alpha=0.2$ in $\R_1$ and
$\R_2$. This difference is useful to model the fact that $\Sc_1$ and $\Sc_2$ most probably interact where they live, rather than {\em on the
way}.

All these figures share a common feature: they all show that $\Sc_1$ leaves $\R_1$, moving towards $\R_2$, while only a small part of $\Sc_2$
moves towards $\R_1$. This is related to the value of the parameters $\gamma_a$ and $\gamma_b$, as well as the $p_{\alpha,\beta}$ which were
fixed at the very beginning, accounting for the diffusion in the model, see (\ref{33}). Since $\gamma_a>\gamma_b$, it is clear that $\Sc_1$ has
a larger mobility than $\Sc_2$. This is exactly what all the figures show. Figures 2(b) and 3(b) show that, when the density of $\Sc_1$ in
$\R_2$ approaches that of $\Sc_2$, $\Sc_2$ reacts very fast in two ways:  their birth rate increases very fast (since its density increases),
and they start rejecting somehow the members of $\Sc_1$ (since the density of $\Sc_1$ decreases). After this first reaction, we  see that, from
time to time, a certain amount of people of $\Sc_1$ goes back to $\R_1$ (presumably, after reaching some well-being). We see that in $\R_2$ the
density of $\Sc_2$ stays almost always larger than that of $\Sc_1$, while in $\R_1$ the density of $\Sc_2$ is always very low: rich people do
not go in the poor area! Moreover, a lot of people of both populations are in $\R_3$: they travel, not necessarily moving from $\R_1$ to $\R_2$
or vice--versa. Incidentally we observe that, because of our interpretation in terms of richness of populations, the parameters $\gamma_a$ and
$\gamma_b$, which are directly proportional to the mobility of the species, can also be seen as inversely proportional to their richness: the
larger the value of $\gamma_a$, the poorer the species, and, consequently, the larger the will to go away from the related cell!

The plots also suggest that the $\omega$'s measure a sort of {\em inertia} of the two populations: increasing the value of, say,
$\omega_b$, produces a less oscillatory behavior of $\Sc_2$, as we can see from Figures \ref{figA1}--\ref{figA3}. Analogously, we have checked
that increasing the value of $\omega_a$, produces a more static behavior of $\Sc_1$. We should also mention that our numerical computations for
$L>11$ confirm our conclusions, showing that the size of $\R$ is not important, except for the time needed to move from $\R_1$ to $\R_2$, as it is
natural. Figures \ref{figA4}--\ref{figA6} show much faster oscillations in the densities of $\Sc_1$ and $\Sc_2$ than those in Figures
\ref{figA1}--\ref{figA3}, in particular in the regions $\R_1$ and $\R_2$. This is due to the fact  that, in these regions, the interaction
parameters between the populations, $\lambda_\alpha$, are taken larger than before. Hence, their densities can change faster than before: the
interaction between $\Sc_1$ and $\Sc_2$ is more important than the diffusion of the populations!

It should also be stressed that the somehow oscillatory behavior which is observed in many figures (and which can also be tested taking a
larger time interval), reflects what observed in \cite{ff2} in a different context, and is, in a sense, intrinsically related to the fact that
we are dealing with a closed physical system ruled by linear ordinary differential equations possessing quasi-periodic solutions. The way our
approach can be extended to include real damping effects is widely discussed in \cite{bag5}, and is based on the introduction of a suitable
{\em reservoir} interacting with the original system. In this case it is possible to check that the time dependence of the number operator
$a_\alpha^\dagger(t) a_\alpha(t)$ can be written as the product of an (essentially) oscillating self-adjoint operator $x_\alpha^\dagger(t)
x_\alpha(t)$, times a decaying factor:
$$
a_\alpha^\dagger(t) a_\alpha(t)=e^{-\frac{2\pi\gamma^2}{\Omega}\,t}\,x_\alpha^\dagger(t) x_\alpha(t),
$$
where $\Omega$ is a parameter of the hamiltonian of the reservoir, see \cite{bag5} for a full analysis in a different context.

\section{Competition between populations}
\label{sec:spreading2}

The same hamiltonian $H$ introduced in Section 3 can be used in the description of competition between two populations $\Sc_1$ and $\Sc_2$.
The equations for the two populations are again those in (\ref{34}). The values of the $p_{\alpha,\beta}$ are chosen exactly as in the previous
migration model: a component of $\Sc_1$ or $\Sc_2$ can move from one cell to a neightboring cell. Hence, the  equation of motion can be written
as in (\ref{310}), and the solution  is given in (\ref{311}). The difference consists here in the choice of the parameters and of the initial
conditions. In fact, in this case, we are no longer necessarily interested in having, at $t=0$, the two populations localized in different
regions of $\R$. Therefore, we consider here again a square lattice, with $L=11$, in which both $\Sc_1$ and $\Sc_2$ are localized in a central
region $\R_c$ of three by three cells, so that they are forced to interact between them from the very beginning.

In Figure \ref{figA7}, in each row, we plot the local densities of $\Sc_1$ (solid line) and $\Sc_2$ (dashed line) inside (left) and outside
(right) $\R_c$. Different rows correspond to different values of $\omega_\alpha^b$, while all the other parameters coincide. We have chosen two
significantly different values of $\gamma_a$ and $\gamma_b$ to give the two populations different mobilities: since $\gamma_a=0.1\gg
\gamma_b=0.004$, $\Sc_1$ is expected to move much faster than $\Sc_2$, and this is exactly what we observe in the figure. Moreover, we have
already seen that $\omega_\alpha^a$ and $\omega_\alpha^b$ play  the role of inertia of the populations in the different cells. Hence, we expect
that the higher is the ratio $\frac{\omega_\alpha^a}{\omega_\alpha^b}$, the smaller will be the reaction of $\Sc_1$ compared with that of
$\Sc_2$. These features are all evident in Figure \ref{figA7}: $\Sc_1$ tends to move away from $\R_c$ faster (or even much faster) than
$\Sc_2$. Moreover, going from the first row ($\omega^a_\alpha=1$, $\omega^b_\alpha=0.3$, $\forall \alpha\in\R$) to the last one
($\omega^a_\alpha=1$, $\omega^b_\alpha=3$, $\forall \alpha\in\R$), it is clear that the tendency of $\Sc_2$ to move away from $\R_c$ decreases
more and more, even if its individuals keep on moving along $\R$.

Particularly interesting is the second row where the density of $\Sc_2$ in $\R_c$ first increases very fast, while that of $\Sc_1$ decreases:
this can be considered as the evidence of a bigger {\em efficiency} of $\Sc_2$ compared with that of $\Sc_1$, which is forced by $\Sc_2$ to
leave $\R_c$.  For instance, thinking of $\Sc_1$ as {\em preys} and of $\Sc_2$ as {\em predators}, we can say that the preys run very fast away
from the region where the predators are localized. Hence $\gamma_a$ and $\gamma_b$ can be considered, other than diffusion coefficients, as a
sort of {\em inverse ability} of the two populations: since $\gamma_b^{-1}\gg\gamma_a^{-1}$, $\Sc_2$ is much stronger than $\Sc_1$, and the
preys are killed significantly by the predators or, if they survive, run away from $\R_c$.

Again, due to the absence of a reservoir, it is not expected any decay for large $t$, and this is exactly what the plots show. In order to have
such a decay, the reservoir must be considered inside the model. This is reasonable, since such a reservoir can play the role of all the
interactions that $\Sc_1$ and $\Sc_2$ may experience other than the mutual interaction (lack of food, other predators, cold winters, hot
summers, \ldots).

\section{Conclusions}
\label{sec:conclusions} In this paper, we have used fermionic operators to describe the dynamical behavior of two populations of individuals
subjected to a certain diffusion. The model considered here is based on a quadratic hamiltonian, so that the resulting equations of motion are
linear and the densities of the populations can be deduced analytically. We have used this rather general hamiltonian in two different
contexts: first, we have described the dynamics of migration of two populations originally spatially separated. In this case we have seen that
the migrants, which are originally well localized in a (poor) region of our lattice, move towards richer zones. This movement is driven not
only from the general form of the hamiltonian but also by the choice of the parameters of the model, which therefore acquire a precise meaning.
The same hamiltonian, with different choices of the parameters and of the initial conditions, has also been used in the description of the
competition between two populations, like in a predator--prey simple system, and we have shown that again a reasonable and interesting dynamics
can be deduced.

Compared with other approaches and results, it is worth to be underlined that the use of fermionic operators automatically ensures the
coexistence of the competing populations in the same environment.

We are aware that the analysis here considered can be seen as a first step towards the construction of a more complete and satisfactory model
of interaction between populations. For instance, damping  and/or nonlinear effects should be taken in consideration. Also, other possible
topologies of the region $\R$ may give interesting results. These are just part of our plans for the future.

\renewcommand{\theequation}{A.\arabic{equation}}

\section*{Appendix:  Few results on the number representation}
We discuss here few important facts in quantum mechanics and in the so--called number representation, paying not much attention to mathematical
problems arising from the fact that the operators involved might be unbounded, since this class of operators is not relevant for the
applications proposed in this paper. More details can be found, for instance, in \cite{mer,rom}, as well as in
\cite{bag1,bag2,bag3,bag4,ff1,ff2,bag5}.

Let $\Hil$ be an Hilbert space, and $B(\Hil)$ the set of all the bounded operators on $\Hil$.    Let $\ST$ be our physical system, and $\A$ the
set of all the operators useful for a complete description of $\ST$, which includes the \emph{observables} of $\ST$. For simplicity, it is
convenient to assume that  $\A$ coincides with $B(\Hil)$ itself. The description of the time evolution of $\ST$ is related to a self--adjoint
operator $H=H^\dagger$ which is called the \emph{Hamiltonian} of $\ST$, and which in standard quantum mechanics represents  the energy of
$\ST$. We will adopt here the so--called \emph{Heisenberg} representation, in which the time evolution of an observable $X\in\A$ is given by
\be X(t)=\exp(iHt)X\exp(-iHt), \label{a1} \en or, equivalently, by the solution of the differential equation \be
\frac{dX(t)}{dt}=i\exp(iHt)[H,X]\exp(-iHt)=i[H,X(t)],\label{a2} \en where $[A,B]:=AB-BA$ is the \emph{commutator} between $A$ and $B$. The time
evolution defined in this way is a one--parameter group of automorphisms of $\A$.

An operator $Z\in\A$ is a \emph{constant of motion} if it commutes with $H$. Indeed, in this case, equation (\ref{a2}) implies that $\dot
Z(t)=0$, so that $Z(t)=Z$ for all $t$.

In our previous papers \cite{bag1,bag2,bag3,bag4,ff1,ff2,bag5}, a special role was played by the so--called \emph{canonical commutation
relations}. Here, these are replaced by the so--called \emph{canonical anti--commutation relations} (CAR): we say that a set of operators
$\{a_\ell,\,a_\ell^\dagger, \ell=1,2,\ldots,L\}$ satisfy the CAR if the conditions \be \{a_\ell,a_n^\dagger\}=\delta_{\ell n}\Id,\hspace{8mm}
\{a_\ell,a_n\}=\{a_\ell^\dagger,a_n^\dagger\}=0 \label{a3} \en hold true for all $\ell,n=1,2,\ldots,L$. Here, $\Id$ is the identity operator
and $\{x,y\}:=xy+yx$ is the {\em anticommutator} of $x$ and $y$. These operators, which are widely analyzed in any textbook about quantum
mechanics (see,  for instance, \cite{mer,rom}) are those which are used to describe $L$ different \emph{modes} of fermions. From these
operators we can construct $\hat n_\ell=a_\ell^\dagger a_\ell$ and $\hat N=\sum_{\ell=1}^L \hat n_\ell$, which are both self--adjoint. In
particular, $\hat n_\ell$ is the \emph{number operator} for the $\ell$--th mode, while $\hat N$ is the \emph{number operator of $\ST$}.
Compared with bosonic operators, the operators introduced here satisfy a very important feature: if we try to square them (or to rise to higher
powers), we simply get zero: for instance, from (\ref{a3}), we have $a_{\ell}^2=0$. This is related to the fact that fermions satisfy the Fermi
exclusion principle \cite{rom}.

The Hilbert space of our system is constructed as follows: we introduce the \emph{vacuum} of the theory, that is a vector $\varphi_{\bf 0}$
which is annihilated by all the operators $a_\ell$: $a_\ell\varphi_{\bf 0}=0$ for all $\ell=1,2,\ldots,L$. Then we act on $\varphi_{\bf 0}$
with the  operators $a_\ell^\dagger$ (but not with higher powers, since these powers are simply zero!): \be
\varphi_{n_1,n_2,\ldots,n_L}:=(a_1^\dagger)^{n_1}(a_2^\dagger)^{n_2}\cdots (a_L^\dagger)^{n_L}\varphi_{\bf 0}, \label{a4} \en $n_\ell=0,1$ for
all $\ell$. These vectors form an orthonormal set and are eigenstates of both $\hat n_\ell$ and $\hat N$: $\hat
n_\ell\varphi_{n_1,n_2,\ldots,n_L}=n_\ell\varphi_{n_1,n_2,\ldots,n_L}$ and $\hat N\varphi_{n_1,n_2,\ldots,n_L}=N\varphi_{n_1,n_2,\ldots,n_L}$,
where $N=\sum_{\ell=1}^Ln_\ell$. Moreover, using the  CAR, we deduce that $\hat
n_\ell\left(a_\ell\varphi_{n_1,n_2,\ldots,n_L}\right)=(n_\ell-1)(a_\ell\varphi_{n_1,n_2,\ldots,n_L})$ and $\hat
n_\ell\left(a_\ell^\dagger\varphi_{n_1,n_2,\ldots,n_L}\right)=(n_\ell+1)(a_l^\dagger\varphi_{n_1,n_2,\ldots,n_L})$, for all $\ell$. The
interpretation does not differ from that for bosons, \cite{bag1}, and then $a_\ell$ and $a_\ell^\dagger$ are again called the
\emph{annihilation} and the \emph{creation} operators. However, in some sense, $a_\ell^\dagger$ is {\bf also} an annihilation operator since,
acting on a state with $n_\ell=1$, we destroy that state.

The Hilbert space $\Hil$ is obtained by taking  the linear span of all these vectors. Of course, $\Hil$ has a finite dimension. In particular,
for just one mode of fermions, $dim(\Hil)=2$. This also implies that, contrarily to what happens for bosons, the fermionic operators are
bounded.

The vector $\varphi_{n_1,n_2,\ldots,n_L}$ in (\ref{a4}) defines a \emph{vector (or number) state } over the algebra $\A$  as \be
\omega_{n_1,n_2,\ldots,n_L}(X)= \langle\varphi_{n_1,n_2,\ldots,n_L},X\varphi_{n_1,n_2,\ldots,n_L}\rangle, \label{a5} \en where
$\langle\,,\,\rangle$ is the scalar product in  $\Hil$. As we have discussed in \cite{bag1,bag2,bag3,bag4,ff1,ff2,bag5}, these states are used
to \emph{project} from quantum to classical dynamics and to fix the initial conditions of the considered system.

\section*{Acknowledgments}

This work has been financially supported in part by G.N.F.M. of I.N.d.A.M., and by local Research Projects of the Universities of Messina and Palermo. The authors acknowledge the unknown referees for the helpful
suggestions that improved the quality of the paper.

\end{document}